\documentclass[preprint2]{aastex}

\renewenvironment{quote} {
  \list{}{
    \leftmargin0.0cm
    \rightmargin\leftmargin
  }
  \item\relax
}
{\endlist}

\shorttitle{Satellite-Mounted Light Sources as Photometric Calibration Standards}
\shortauthors{Albert, J.}

\begin{document}


\title{SATELLITE-MOUNTED LIGHT SOURCES AS PHOTOMETRIC CALIBRATION STANDARDS FOR GROUND-BASED TELESCOPES}


\author{J.~Albert}
\affil{Department of Physics and Astronomy, University of Victoria,
    Victoria, BC V8P 5C2 Canada}
\email{jalbert@uvic.ca}



\begin{abstract}
A significant and growing portion of systematic error on a number of fundamental parameters in astrophysics and
cosmology is due to uncertainties from absolute photometric and flux standards.  A path toward achieving major
reduction in such uncertainties may be provided by satellite-mounted light sources, resulting in improvement in the
ability to precisely characterize atmospheric extinction, and thus helping to usher in the coming generation of
precision results in astronomy.  Using a campaign of observations of the 532 nm pulsed laser aboard the CALIPSO
satellite, collected using a portable network of cameras and photodiodes, we obtain initial measurements of
atmospheric extinction, which can apparently be greatly improved by further data of this type.  
For a future satellite-mounted precision light source, a high-altitude balloon platform 
under development (together with colleagues) can provide testing as well as observational data for calibration of atmospheric 
uncertainties.
\end{abstract}


\keywords{techniques: photometric, instrumentation: photometers, atmospheric effects, dark energy, space vehicles: instruments, balloons}



\section{INTRODUCTION}

While our understanding
of the Universe has changed and improved dramatically over the past 25 years, the improvement of our knowledge
of absolute spectra and flux from standard calibration sources, upon which the precision of measurements of
the expansion history of the Universe \citep{alb06}, and of stellar and galactic evolution [\textit{e.g.}~\citet{eis01}] are based, 
has been far
slower and not kept up with reduction of other major uncertainties.  As a result, uncertainties on absolute
standards now constitute one of the dominant systematics for measurements such as the expansion history
of the Universe using type Ia supernovae [\textit{e.g.}~\citet{woo07}, \citet{ast06}, \citet{kno03}], and a significant systematic 
for
measurements of stellar population in galaxy cluster counts \citep{ken09,koe07}, and
upcoming photometric redshift surveys measuring growth of structure \citep{con06}.  There
are prospects for improvement in uncertainties from standard star flux and spectra \citep{kai08}, but the
traditional techniques of measurement of standard stellar flux from above the atmosphere suffer from basic
and inherant problems: the variability of all stellar sources, and the difficulty of creating a precisely calibrated,
cross-checked, and stable platform for observation above the Earth's atmosphere.

The presence of an absolute flux standard in orbit above the Earth's atmosphere could provide important
cross-checks and potential significant reduction of photometric and other atmospheric uncertainties for
measurements that depend on such calibration.  Monochromatic sources, especially ones which could cover
multiple discrete wavelengths, or tune over a spectrum, could potentially help to further reduce spectrophotometric error.
For calibration of telescope optics and detector characteristics, authors \citep{stu06} have both conceived of and used a
wavelength-tunable laser within present and upcoming telescope domes as a color calibration standard.
Although a wavelength-tunable laser calibration source in orbit \citep{alb06j} does not exist yet, at present there is a
532 nm laser in low-Earth orbit pointed toward the Earth's surface, with precise radiometric measurement of
the energy of each of the 20.25 Hz laser pulses, on the CALIPSO satellite, launched in April 2006 \citep{win09}.  We have
collected data from a portable network of seven cameras and two calibrated photodiodes, taken during CALIPSO flyovers
on clear days in various locations in western North America.  The cameras and photodiodes respectively capture images and
pulses from the eye-visible green laser spot at the zenith during the moment of a flyover.  Using precise pulse-by-pulse
radiometry data from the CALIPSO satellite, we compare the pulse energy received on the ground with the pulse energy
recorded by CALIPSO.  The ratio determines the atmospheric extinction.  

These measurements apply only at the zenith, at the specific location of the telescope network, for the
specific laser frequency, at the time of the CALIPSO flyover.
To obtain results that are more easily applicable to modern astronomy, a satellite laser source must be able to
point at major observing facilities, and modify its frequency to be able to scan through the visible and near-infrared spectrum.  
CALIPSO does not have this ability, however future atmospheric science satellite missions may have such capabilities.
A brief discussion of this is in Section 8.

For the analysis of the images and the determinations of absolute flux, both laboratory and field calibration of the cameras
and the photodiodes was essential.  We developed a calibration system to determine the sizes of effects such as the
anisotropy, nonlinearity, and temperature dependence of the responses of the CCDs, so that we could accurately and 
precisely measure optical energy of a
source from its image.  We were able to calibrate the cameras to a measured uncertainty of 2.5\% or better in absolute
flux.  We include a detailed discussion of systematic effects and uncertainties on our measurements, currently dominated
by atmospheric scintillation.  Following the discussion of
our results, we elucidate ways that this technique could be considerably improved beyond this initial study.

\section{SATELLITE-MOUNTED LIGHT \\ SOURCES}
\label{satellites}

Throughout history prior to 1957, the only sources of light above the Earth's atmosphere were natural in origin: stars, and 
reflected light from planets, moons, comets, etc.  Natural sources have of course served extremely well in astronomy: through 
understanding the physical processes governing stellar evolution, we are now able to precisely understand the spectra of stars used 
as calibration sources [\textit{e.g.}~\citet{boh00}].  Nevertheless, in all stars the vast bulk of material, and the thermonuclear processes that themselves 
provide the light, lie beyond our sight below the surface of the star.  Superb models of stellar structure are available, but 
uncertainties of many types always remain.

Since the launching of the first man-made satellites, a separate class of potential light sources in space has become available. 
Observable light from most satellites is primarily due to direct solar reflection, or reflection from Earth's albedo.  While 
providing a convenient method of observing satellites, this light is typically unsuitable for use as a calibrated light source due 
to large uncertainties in the reflectivity (and, to a lesser extent, the precise orientation and reflective area) of satellites' 
surfaces. Reflected solar light has, however, been successfully used as an absolute infrared calibration source by the Midcourse 
Space Experiment (MSX), using 2 cm diameter black-coated spheres ejected from the MSX satellite, whose infrared emission was 
monitored by the instruments aboard MSX~\citep{pri04}.  This technique proved highly effective for the MSX infrared calibration; 
however, the technique is not easily applicable to measuring extinction of visible light in the atmosphere.

Many satellites have retroreflective cubes intended for use in satellite laser ranging.  Reflected laser light from retroreflectors 
is critical for distance measurements using precise timing; however, like solar reflection, retroreflected laser light 
unfortunately also suffers from uncertainties in the reflectivity of the cubes, and in reflectivity as a function of incident 
angle, that are too large to provide a means of measuring atmospheric extinction \citep{min74}.  
A satellite-based mirror, either planar or spherical, which could reflect either sunlight or a ground-based light source, 
would also similarly require some means of calibrating reflectivity, both as a function of incidence angle and as a function of time (due
to pitting from dust particles intersecting its orbit)---although it is a potentially interesting concept.
Thus we are left with dedicated 
light sources aboard satellites themselves as the practical means of having a satellite-based visible light source for 
precise calibration of ground-based telescopes.

Many satellites also carry some means of producing observable visible light, for self-calibration purposes or otherwise.  The Hubble 
Space Telescope is one of many satellites carrying tungsten, as well as deuterium, lamps for absolute self-calibration purposes 
\citep{pav01}.  Lamps for self-calibration are not limited to space telescopes for astronomy; earth observation and weather 
satellites also commonly use internal tungsten lamps as calibration sources \citep{nit93}. Such internal calibration lamps are 
typically limited by the fact that they can degrade individually, and can be compared only with astronomical sources after launch, 
leaving stellar light as the only practical way to ``calibrate the calibration device.'' Thus, such devices typically provide a 
cross-check rather than the basis for a true absolute irradiance calibration, or provide a means for a separate calibration, such 
as flat-field [\textit{e.g.}~\citet{boh04}].  Furthermore, present-day internal calibration lamps aboard satellites are certainly 
not intended for, nor are capable of, a direct calibration of the atmospheric extinction that affects ground-based telescopes.

However, a satellite-based absolute calibration source for ground-based telescopes is not technically prohibitive.
As an example, a standard household 25-watt tungsten filament lightbulb (which typically have a temperature of the order of
3000 K and usually produce approximately 1 watt of visible light between 390 and 780 nm) which
radiates light equally in all directions from a 700 km low Earth orbit has an equivalent brightness to a 12.5-magnitude star
(in the AB system, although for this approximate value the system makes little difference).  In
general, the apparent magnitude of an orbiting lamp at a typical incandescent temperature which radiates isotropically is 
approximately given by
\begin{equation}
m \approx -5.0 \log_{10} \left ( \frac{ \left ( \ln \left ( \frac {P}{\rm{1\;watt}} \right ) \right )^3 }{h} \right ) + 5.9,  
\end{equation}
where $P$ is the power of the lamp in watts, and $h$ is the height of the orbit in kilometers.

The dominant uncertainties in the amount of light received by a ground-based telescope from such an orbiting lamp would 
stem from uncertainties in onboard radiometry monitoring the power output of the lamp,
any unsubtracted background from reflected earthshine, sunshine, moonshine, or starlight from the surface of the 
satellite itself, and potential deviations from perfect isotropic output (i.e.~differences in the onboard-monitored output vs.~the 
ground-observed output) of the light from the lamp. 
The uncertainty on the magnitude of the lamp stemming from uncertainties in the radiometrically-monitored output
power would be limited by the precision of current radiometer technology.  Modern radiometers, using electrical substitution 
radiometry, can achieve a precision of approximately 200 parts per million, with the dominant uncertainty being the size of
the aperture~\citep{kop05}.  
The uncertainty on 
the magnitude of the lamp due to unsubtracted background from reflected earthshine, sunshine, moonshine, or starlight from the surface of the 
satellite would depend both on the size and surface material of the satellite, and on the performance of shuttering of the lamp
to provide images to subtract such background.  To minimize reflectance, one could coat the surface of the satellite with 
a black nonreflective coating, however this would result in heating of the satellite as it passes through sunlight, until such 
heating was in equilibrium with thermal radiation from the satellite, an average temperature of approximately 280 K for a spherical 
nonreflective satellite in continuous sunlight that generates no internal power, and 280 K $\times \left( 1 + 9.3 \times 10^{-4} 
\left( \frac{P}{d^2} \right) \right )^{1/4}$ for such a spherical nonreflective satellite of diameter 
$d$ meters that generates $P$ watts of internal power.  The temperature differences as the satellite passes through sunlight and 
through the Earth's shadow would be quite large, and power supplies for a lamp (and the lamp itself) would need to allow for such 
an operating temperature shift.  Assuming this is achieved, there would be residual reflectance from the black coating.  Typical 
NiP black coatings can achieve a reflectance of as low as 0.2\% throughout most of the visible range~\citep{kod90}.
Assuming perfect shuttering of the lamp output, the image-to-image variance in reflected light as the satellite passes through Earth's shadow would 
form the uncertainty, a value that is likely small and could be calibrated.  
Finally, the uncertainty on the magnitude of the lamp 
due to potential deviations from perfect isotropic output of the light from the lamp would be able to be tightly constrained using 
measurements done in the laboratory before launch, as well as minimized by placing the lamp inside an integrating sphere, and thus should be a very 
small, if not a negligible, contribution.  Remaining 
uncertainties from this source could stem from uncertainty in the precise orientation of the satellite, and/or effects from 
degradation of the lamp on the isotropic output of the light, but would likely be small.  Thus the predicted total sysmatic 
uncertainty on the radiance of an optimally-designed orbiting lamp would be dominated by the precision of
radiometric monitoring technology, and be in the range of approximately 200 parts per million if the best presently available 
radiometric technology is used.

An alternative to an isotropic or near-isotropic lamp would be a laser source, with beam pointed at the observer (with a small 
moveable mirror, for example).  Divergences of laser beams are typically on the order of a milliradian (which can be reduced to 
microradians with a beam expander) so much less output power than a lamp would be required for a laser beam to mimic the brightness 
of a typical star. (However, note that the wall-plug power of typical diode-pumped lasers is typically in the vicinity of $20$ 
times the output laser power [\textit{e.g.}~\citet{sea07}], which is furthermore a large reduction in wall-plug power from the 
$10^{4}$ level typical of flashlamp-pumped lasers.) The apparent magnitude of an orbiting laser with Gaussian beam divergence 
pointed directly at a ground-based telescope, is given by
\begin{equation}
m \approx -2.5 \log_{10}\left ( \frac{P}{h^2 d^2} \right ) - 20.1,
\end{equation}
where $P$ is the laser power in milliwatts, $h$ is the
height of the orbit in kilometers, and $d$ is the RMS divergence of the laser beam in milliradians,
under the assumption that the aperture of the telescope is small compared with the RMS width of the beam at the ground, $hd$.
The RMS divergence would be the combination of the divergence at the source, and the divergence due to the atmosphere.  In
clear conditions, total atmospheric divergence in a vertical path is at the level of approximately 5 microradians~\citep{tat61}, 
and this of course only
acts on the last fraction of the laser path that is within the atmosphere, so as long as the source divergence is
significantly larger than this, atmospheric divergence would be negligible. 
Clearly, even a sub-milliwatt laser in low Earth orbit would need to have either its divergence increased at the source, or its power reduced
via filtering, for it to be suitable for astronomical calibration.  We shall consider the filtering option.
The major uncertainties in the amount of light received
by a ground-based telecope from an orbiting laser would stem from uncertainties in the pointing and beam profile of the
laser light (which would likely need to be monitored by an array of small dedicated telescopes outboard of the main
ground-based telescope), in time-dependent variation of the laser output power (which would likely need to
be monitored by onboard radiometry), and in degradation of the filter over time.  Nevertheless, laser light has the benefit
of being monochromatic, allowing for calibration of individual wavelengths.  With a widely-tunable laser, an entire spectrum
could be calibrated, removing the significant inherant uncertainties associated with comparing the spectra of astrophysical objects 
with the spectrum of a calibration source.

The uncertainty on the apparent magnitude of an orbiting laser stemming from uncertainties in the radiometrically-monitored laser 
power would be limited by the precision of current radiometer technology.  Modern electrical substitution 
radiometers can achieve a precision of approximately 100 parts per million when aperture uncertainties can be neglected, as in the
case of laser radiometry~\citep{kop05}.  
Uncertainties on the magnitude due to 
uncertainty in the pointing and beam profile would potentially be limited by the size of the array of outboard telescopes for 
monitoring the laser spot, and by calibration differences between the individual telecopes in the array and with the main central 
telescope.  The latter could clearly be minimized by a ground system for ensuring the relative calibration of the outboard 
telescopes and main telescope are all consistent.  Thus one could achieve a similar, or potentially even smaller, uncertainty on
apparent magnitude from an orbiting laser source as from an orbiting lamp source, but with the major added complication of the need
for a large, dense array of outboard telescopes to very precisely monitor the position of the centroid of the laser beam on the ground.

The uncertainties considered above assume that the exposure time is long compared with the coherence time of the atmosphere.  With 
short exposures, or in the case of a laser that either quickly sweeps past, or is pulsed, atmospheric scintillation can play a 
major role in uncertainty in apparent magnitude of a satellite-mounted source.  A typical timescale for a CW laser with 1 
milliradian divergence in low Earth orbit to sweep past is tens of milliseconds, which is of the same order as characteristic 
timescales of atmospheric scintillation, and the typical timescale of single laser pulses is nanoseconds, much shorter than 
scintillation timescales, thus one cannot assume that such effects can be time-averaged over.  In idealized conditions, for small 
apertures $D < \sim 5$ cm and sub-millisecond integration times, the relative standard deviation in intensity $\sigma_I \equiv 
\Delta I / \langle I \rangle$, where $\Delta I$ is the root-mean-square value of $I$, is given by the square root of
\begin{equation}
\sigma_I^2 = 19.12 \lambda^{-7/6} \int_0^\infty C_n^2(h)h^{5/6}dh,
\end{equation}
where $\lambda$ is optical wavelength (in meters), $C_n^2(h)$ is known as the refractive-index structure coefficient, and $h$ is 
altitude (in meters) \citep{tat61}.  Large apertures $D > \sim 50$ cm have a relative standard deviation in intensity given by the square root of
\begin{equation}
\sigma_I^2 = 29.48 D^{-7/3} \int_0^\infty C_n^2(h)h^{2}dh
\end{equation}
\citep{tat61}. The values and functional form of $C_n^2(h)$ are entirely dependent on the particular atmospheric conditions at the 
time of observation, however a relatively typical profile is given by the Hufnagel-Valley form:
\begin{eqnarray}
\!\!\! C_n^2(h) \! & \!\!\! = \!\!\! & \! 5.94 \times 10^{-53}(v/27)^2 h^{10}e^{-h/1000} + \nonumber\\
                \! &                 & \! 2.7 \times 10^{-16} e^{-h/1500} + Ae^{-h/100},
\end{eqnarray}
where $A$ and $v$ are free parameters \citep{huf74}.  Commonly-used values for the $A$ and $v$ parameters, which represent the 
strength of turbulence near ground level and the high-altitude wind speed respectively, are $A = 1.7 \times 10^{-14}$ m$^{-2/3}$ 
and $v = 21$ m/s \citep{rog96}. Using these particular values, for a small aperture, the relative standard deviation $\sigma_I$ would be 
expected to be 0.466 for 532 nm light, which is not far off experimental scintillation values for a clear night at a 
typical location [\textit{e.g.}~\citet{jak78}]. For a single small camera, this is an extremely large uncertainty.  Other than by 
increasing integration time (which is not possible with a pulsed laser) or by significantly increasing the camera aperture, the 
only way to reduce this uncertainty is to increase the number of cameras.  With $N$ cameras performing an observation, which are 
spaced further apart than the coherence length of atmospheric turbulence (typically 5 to 50 cm), the uncertainty from scintillation 
can be reduced by a factor $\sqrt{N}$ (for large $N$).

The analysis above considers a hypothetical pointable satellite-mounted calibration laser, and is necessarily both speculative and 
approximate.  However, at present there is an actual laser in low Earth orbit, visible with both equipment and with the naked eye, 
and analysis of ground-based observational data of the laser spot can be used for comparisons with the above, as well as for 
development of and predictions for potential future satellite-based photometric calibration sources of ground telescopes.

\begin{figure*}[t]
\begin{center}
\includegraphics[angle=270,width=16cm]{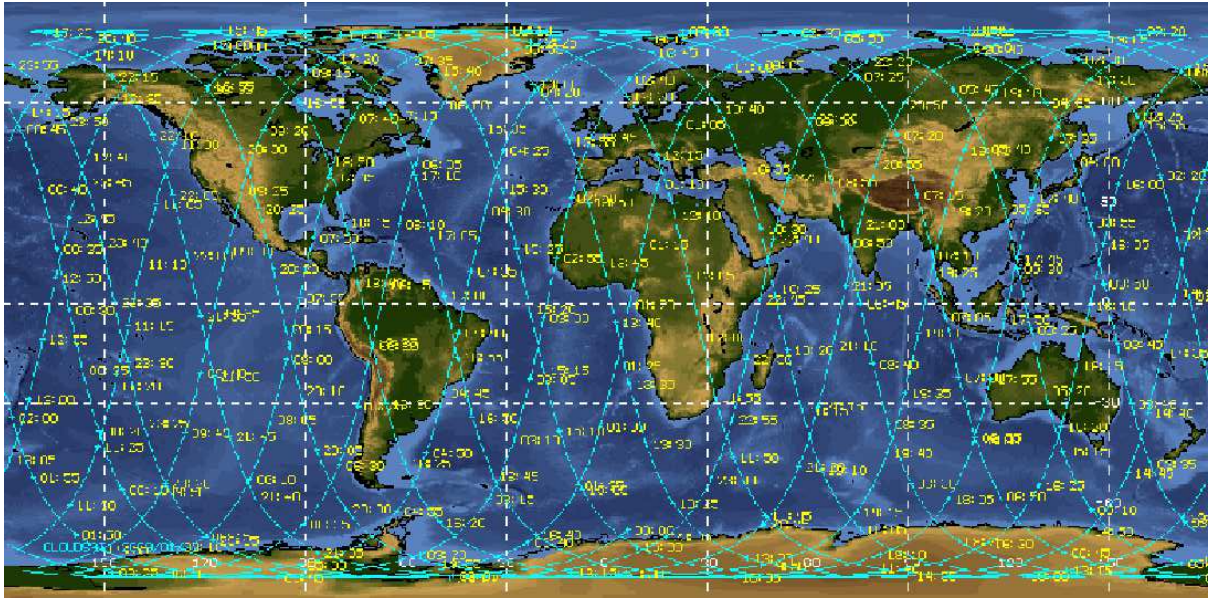}
\caption{
A typical 1-day ground track of the CALIPSO satellite~\citep{gar07}.
}
\label{fig:groundtrack}
\end{center}
\end{figure*}

\section{THE CALIPSO SATELLITE AND GROUND-BASED OBSERVATION NETWORK (2006-08)}
\label{hardware0608}

The CALIPSO (Cloud Aerosol Lidar and Infrared Pathfinder Satellite Observations) satellite was launched on April 28, 2006 as a 
joint NASA and CNES mission~\citep{win09}.  CALIPSO is part of a train of seven satellites (five of which are orbiting at the date 
of this article), known as the ``A-Train,'' in sun-synchronous orbit at a mean altitude of approximately 690 km~\citep{sav08}. A 
typical 1-day ground track for CALIPSO is shown in Fig.~\ref{fig:groundtrack}; CALIPSO completes an orbit every 98.4 minutes 
(approximately 14.6 orbits per day), and repeats its track every 16 days.  CALIPSO contains a LIDAR (Light Detection and Ranging) 
system, known as CALIOP (Cloud Aerosol Lidar with Orthogonal Polarization), with a primary mission of obtaining high resolution 
vertical profiles of clouds and aerosols in the Earth's atmosphere~\citep{hun09}.  The CALIOP laser produces simultaneous, 
co-aligned 20 ns pulses of 532 nm and 1064 nm light, pointed a small angle (0.3$^\circ$) away from the geodetic nadir in the 
forward along-track direction, at a repetition rate of 20.16 Hz.  The light enters a beam expander, following which the divergence 
of each laser beam wavelength is approximately 100 $\mu$rad, producing a Gaussian spot of approximately 70 m RMS diameter on the ground.  
The pulse energy is monitored onboard the satellite, and averages approximately 110 mJ, at each one of the two wavelengths, per 
pulse.  The effective apparent magnitude of the 532 nm laser spot at the precise center of the beam is thus approximately -19.2,
however this high brightness, of course, falls off rapidly as one moves away from the center of the beam.

\begin{figure*}[t]
\begin{center}
\includegraphics[angle=0,width=16cm]{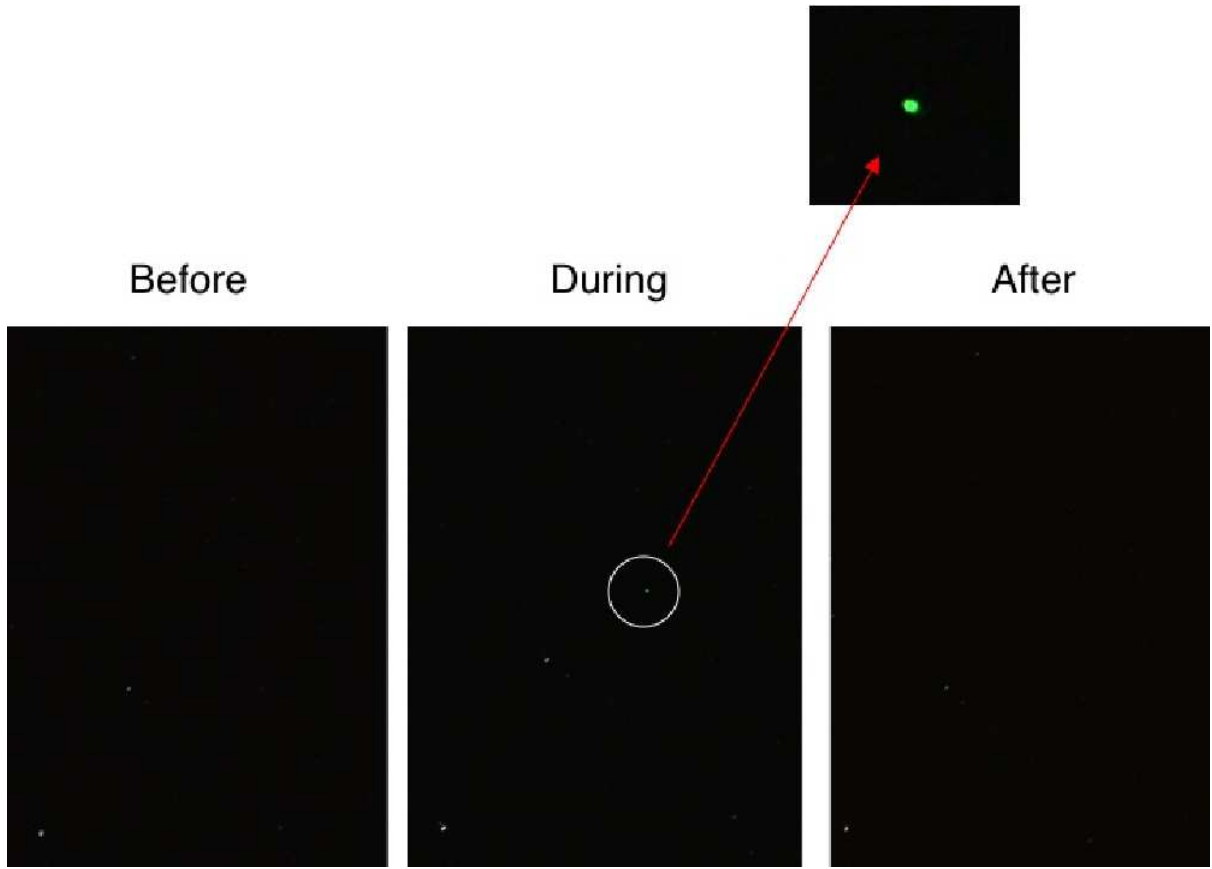}
\caption{
Initial single-camera ground observation of a CALIPSO laser pulse, taken at 07:08:30 UTC on Nov.~23, 2006 near Granby, Quebec.
}
\label{fig:initialobs}
\end{center}
\end{figure*}

\begin{figure*}[t]
\begin{center}
\includegraphics[angle=0,width=16cm]{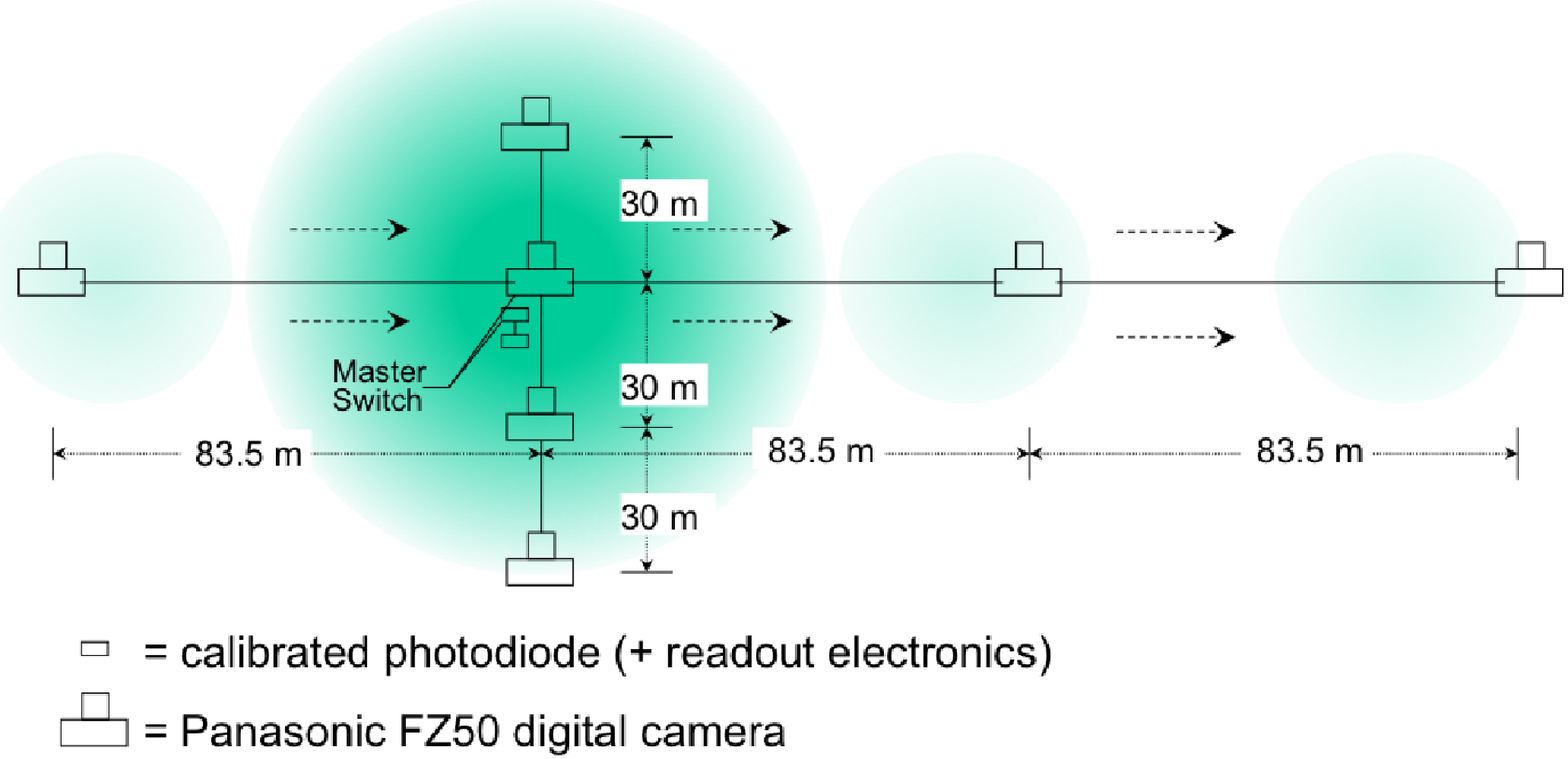}
\caption{
CALIPSO satellite observation setup using seven Panasonic FZ50 cameras connected by coaxial cable, to obtain images of four pulses 
of the CALIPSO laser, and obtain constraints on the energy, position, and width of the beam pulse(s) on the ground.
}
\label{fig:groundstation}
\end{center}
\end{figure*}

During 2007, the CALIPSO beam was observed at several locations in western North America using a portable ground station 
consisting of seven simultaneously-shuttered Panasonic FZ50 digital cameras, connected as shown in Fig.~\ref{fig:groundstation}, 
and two calibrated photodiodes co-located with the central camera.  
One of the two photodiodes is a Hamamatsu S2281 silicon photodiode supplied
by the National Institute of Standards and Technology (NIST, in Gaithersburg, MD), and calibrated by NIST to an absolute radiometric standard
in the visible and near-infrared.  A 12.5 mm diameter Semrock 532 nm line filter was mounted over the front face of this photodiode.
The other photodiode is a Hamamatsu S1336-5BQ, with no filter.  

\begin{figure*}[t]
\begin{center}
\includegraphics[angle=0,width=16cm]{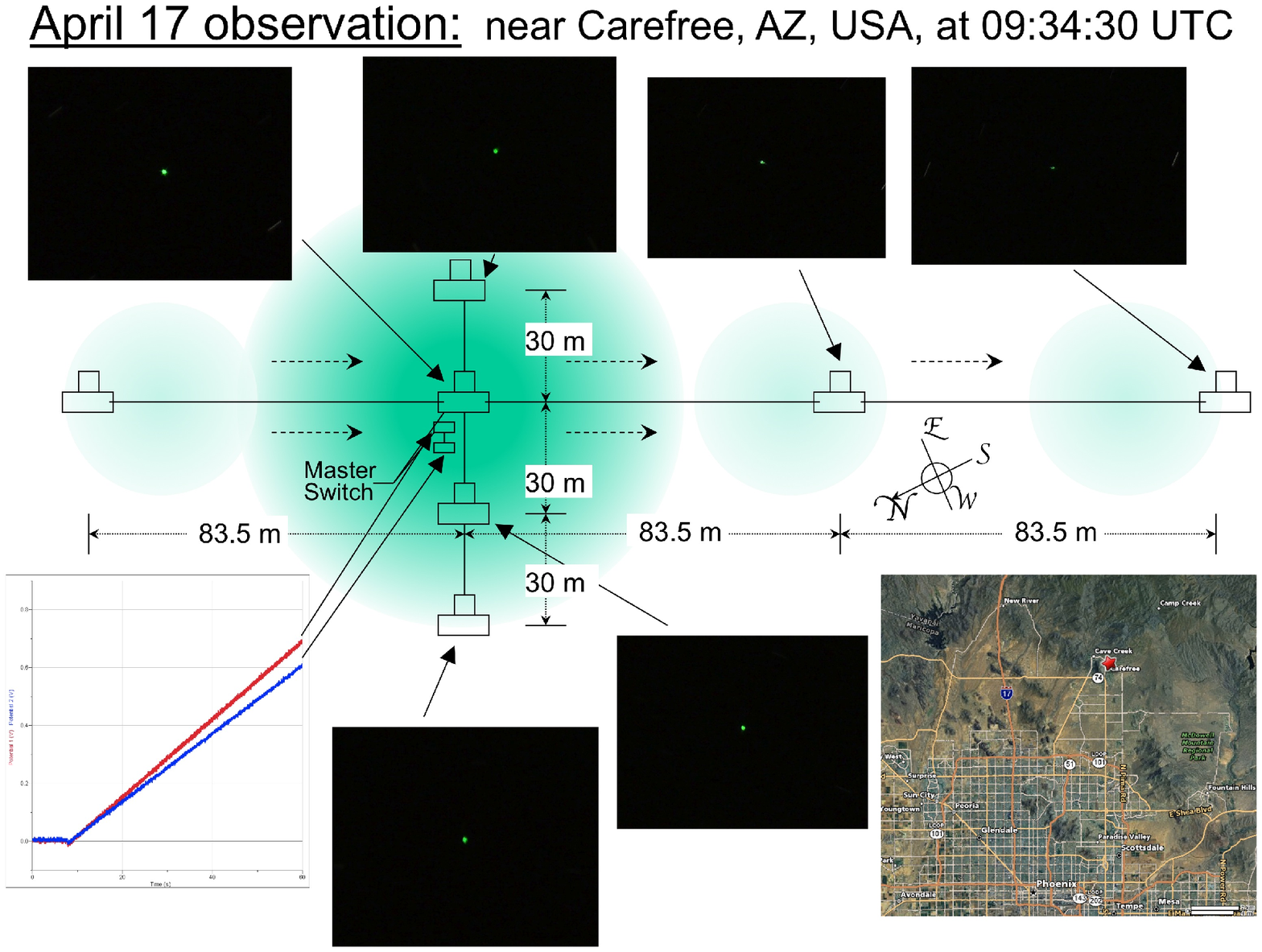}
\caption{
Seven-camera observation of a CALIPSO overpass, taken on Apr.~17, 2007 near Carefree, Arizona.
}
\label{fig:Carefreeobs}
\end{center}
\end{figure*}

\begin{figure*}[t]
\begin{center}
\includegraphics[angle=0,width=16cm]{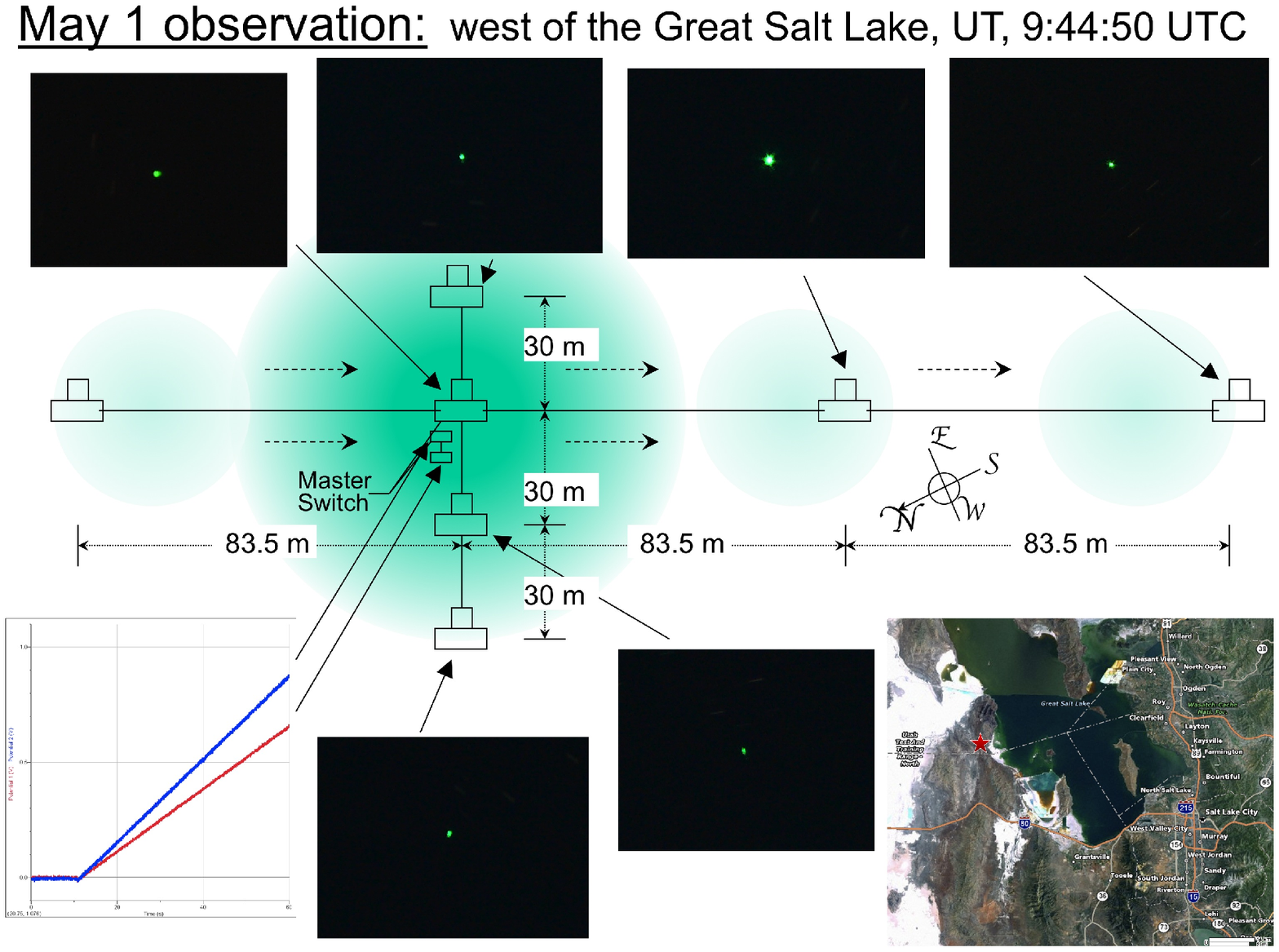}
\caption{
Seven-camera observation of a CALIPSO overpass, taken on May 1, 2007 near the Great Salt Lake, Utah.
}
\label{fig:SaltLakeobs}
\end{center}
\end{figure*}

A list of the observations can be found in 
Table~\ref{tab:observations}.  Two example observations can be seen in Figs.~\ref{fig:Carefreeobs} and \ref{fig:SaltLakeobs}.
Observing locations were selected by means of the CALIPSO ground track, monitored by NASA.  For each 
observation, a point along the ground track was selected the day prior to the night of the overpass by means of the following 
criteria: road accessibility, lack of obscuring trees and streetlights, avoidance of fenced-off private land, and local weather 
conditions. A desert environment is ideal; for this reason, several of the observations were performed in the southwestern U.S.  
Even thin cloud cover can create a difference in time-averaged optical density between the CALIPSO beam path and light from stellar 
standard sources on the image nearby; thus all observations were performed on clear nights.

Each camera was placed pointing vertically, with the images approximately centered on the zenith, for all observations. 
Tripods were used for all observations except for the first (Apr.~17), in order to reduce issues of dust, or condensation,
obscuring the camera lenses (for the Apr.~17 observation, the cameras were placed directly on the ground).  The exposure time 
for all observations with all cameras is 60 s., allowing sufficient exposure to capture images of nearby stars as well as the 
CALIPSO beam.  For each camera, the aperture f-stops were set at f/4.0.  As the lenses are 55 mm focal length, the aperture
diameters were thus 13.75 mm.  The images sensors for Panasonic FZ50 cameras are 1/1.8'' optical format CCDs 
(5.319 mm x 7.176 mm) with 10.1 million effective pixels.
The field of view of each camera is 20.5 x 15.5 degrees; the axes are randomly oriented with respect to RA and 
declination. The two photodiodes, co-located with the central camera, were also pointed toward the zenith and each surrounded by a 
blackened 1 cm radius tube, extending 10 cm above the face of the photodiode, in order to reduce background light.  Approximately 90 
minutes prior to each overpass, a test image was taken with all seven cameras plus the two photodiodes, to ensure all observing 
elements were properly functioning.  CALIPSO beam exposures were started approximately 30 seconds before the overpasses.  About 
half of all observation attempts were unsuccessful in obtaining an image of the CALIPSO beam in all cameras.  The most common 
reason for failed attempts was uncertainty in the CALIPSO ground track.  The uncertainty for CALIPSO ground track is approximately 100 
meters, however this is the uncertainty of the \textit{a postiori} CALIPSO ground track data (available three days after it is taken);
for the \textit{a priori} ground
track forecast, the uncertainty is significantly greater, and is often subject to systematic biases.  Thus there is a very large 
chance of missing the beam.
In addition, in the few days following drag makeup maneuvers by the CALIPSO satellite (which are 
performed approximately once per month), the ground track forecast is especially subject to large systematic offsets.
The second most common cause of failed observations was condensation, or frost, on camera lenses; this was also a difficult problem 
to ameliorate for cameras with lenses that are neither shielded nor heated, other than by only choosing sites in which humidity is very low.

\section{CAMERA AND PHOTODIODE DATA FROM 2006-08}
\label{data}

The data taken by the Panasonic cameras is recorded in a proprietary Panasonic format (referred to as a raw image).  After transferring the 
data to computer disk, the raw images were converted into 16-bit TIFF images (using the sRGB colorspace) using the Silkypix 
Developer Studio 2.0 SE program (v.~2.0.14.10).  Then, using the GIMP 2.0 program, the TIFF images were separated into their red, 
green, and blue components, and each component color in each image was converted to a FITS image file.  Using the DS9 program 
(v.~5.51), the approximate pixel location of the centroid of the laser spot on each image containing the green pixel data was 
determined manually.  Using SuperMongo (v.~2.4.34), the pixel values in a 51 pixel by 51 pixel square centered around the approximate 
centroid pixel location were written out to a text file.

The laser spots are easily found on the original color images by eye-scanning for a green-colored dot that has no trail.  (The
minute-long exposures are enough to create visible trails on stars and other celestial objects, whereas the laser flash, being
essentially instantaneous, creates no trail.)  In all cases where a laser spot existed, it was obvious by eye, and there never 
was a significant question of whether it was the laser, or a green-colored celestial object.  In the FITS files corresponding to
the red and blue parts of the images, little or no signal from the CALIPSO spots is found, and thus the red and blue components add
very little information to the measurement of the signal from the CALIPSO beam.  Thus only the green parts of the images are 
used in the final analysis.

The data from the two Hamamatsu photodiodes was amplified by means of a Hamamatsu C9052 Si photodiode readout board in the 
light integration circuit configuration.  
The output was then subtracted for constant background light and dark current contribution, the former of which varied accoring to 
the amount of moonlight, etc., in the background, and thus was adjustible on-site by means of a potentiometer.  The 
background-subtracted signal was then further amplified by a factor of 50.  The signal was then digitized and read out at 200 Hz 
over the 60 s interval to computer disk by means of a LoggerPro$^{\rm TM}$ readout device; then converted to a text file.  Power 
was supplied via an AC inverter and automobile engine.

Only 1 of the 6 camera observations (the one taken on 30 May) has properly-recorded photodiode data.  Photodiode readout 
electronics, or other, 
problems prevented successful recording in the other 5 observations. 
While the photodiode signal was visible in the 30 May observation, we observed significant differences in the ratio of photodiode-to-camera
signal size from that which we measured in the laboratory, which could be due either to background moonlight affecting the photodiode signal, or differences in
the gain of the photodiode electronics from that measured in the laboratory due to the lower temperature.
Thus, we chose not to use the photodiode data which was collected in the field, even that from the 30 May observation, in the analysis.
Nevertheless, as the cameras were calibrated to the NIST photodiode absolute standard in the laboratory, the data from the seven-camera network 
allows the measurement of the CALIPSO pulse energies reaching the ground.

\section{ANALYSIS}
\label{analysis}

\subsection{Absolute Photometric Calibration of the Panasonic Cameras}

The Panasonic cameras were calibrated to an absolute radiometric standard through the use of the NIST-calibrated Hamamatsu photodiode (see above)
and a Pavillion Integration W532-10FS low-noise 532 nm laser with Newport neutral density filters mounted in the beam to reduce the intensity by a 
factor of approximately $2 \times 10^5$.  The cameras were mounted in the intensity-reduced beam, inside a light-tight enclosure, and each exposed 
for 1 minute, as during CALIPSO observations.  The cameras were then swapped with the calibrated photodiode and readings were taken.  The analyzed 
images were then compared with the photodiode data to obtain a radiometric comparison point.

\begin{figure}[t]
\begin{center}
\includegraphics[angle=0,width=8cm]{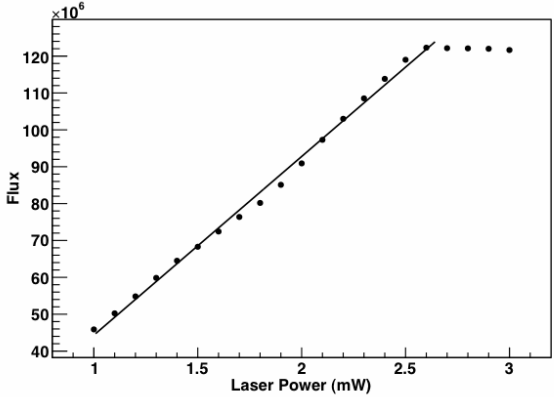}
\caption{
Measurement of the response linearity of the central camera CCD.  The response is linear to within 2.4\% over the energy scale of interest
(\textit{i.e.}~below the plateau due to pixel saturation).  
Note that the laser power is attenuated, by a constant factor of $\mathcal{O}(10^5)$, using neutral density filters.
The response of the other cameras has similar linearity.
}
\label{fig:linearity}
\end{center}
\end{figure}

\begin{figure}[t]
\begin{center}
\includegraphics[angle=0,width=7.5cm]{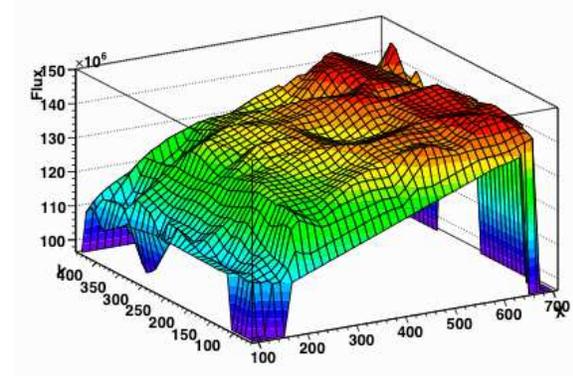}
\caption{
Anisotropy measurement of the CCD of the west camera.  The horizontal axes show the $x$ and $y$ pixel position; the vertical axis is flux, in
arbitrary units.
The deviations from a flat distribution on the vertical axis are magnified by a factor 
of 100.  The other cameras have a similar degree of anisotropy ($\sim$0.5\%).
}
\label{fig:anisotropy}
\end{center}
\end{figure}

\begin{figure}[!b]
\begin{center}
\includegraphics[angle=0,width=7.5cm]{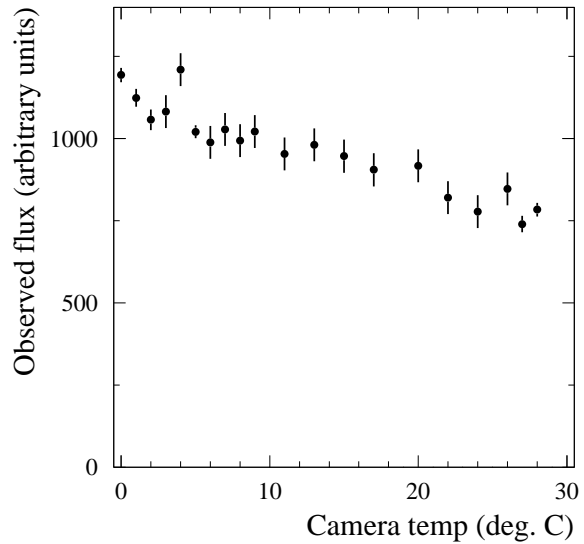}
\caption{
Measurement of the temperature dependence of the observed flux from a stable, calibrated laser source in the central camera (other cameras are similar).
}
\label{fig:tempdependence}
\end{center}
\end{figure}

Measurements of the response linearity, an\-isotropy, and ambient temperature dependence of the cameras were also performed.  Linearity was 
measured by modifying the power of
the laser, using input power control hardware supplied by the laser manufacturer.  The radiometric calibration process above was repeated at 10 different 
laser power settings for two of the cameras.  Linearity was found to be maintained to $\pm2.4\%$, as shown in Fig.~\ref{fig:linearity}.  The anisotropy 
of the response of the CCDs was 
measured by obtaining images as above with the laser spot at 900 different places on each of the cameras' CCDs.  The anisotropy, taken to
be the standard devation of those 900 measurements, was found to 
be 0.5\%, as shown in Fig.~\ref{fig:anisotropy}.  Temperature calibration was performed using one of the cameras inside a sealed refrigerator
with a small hole in it to allow entrance of laser light (from inside a light-tight tube).  Over a temperature range from $20^{\circ}$C down to
$0^{\circ}$C, the image intensity was found to increase by the surprisingly large value of 52\%, as shown in Fig.~\ref{fig:tempdependence}.  
This variation, although large, is not an unexpected phenomenon~\citep{hea94}.

\subsection{Measurement of Camera Positions}

\begin{figure*}[p]
\begin{center}
\includegraphics[angle=0,width=14cm]{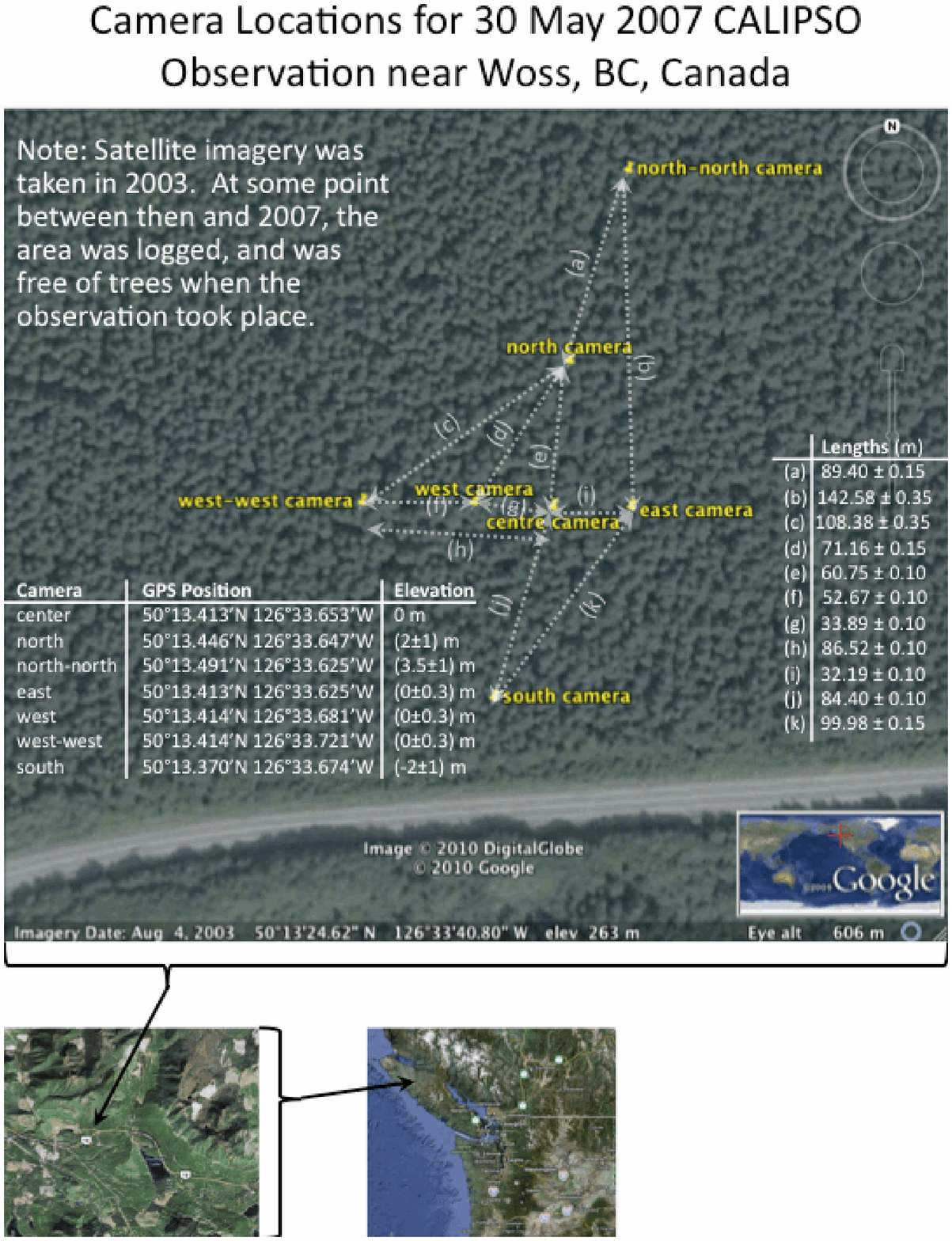}
\caption{
Camera locations for the 30 May 2007 observation (as an example).  Lengths were measured with a surveyor's tape measure, and uncertainties are estimated.
Elevations are relative to the central camera and are estimated by eye (their uncertainties do not contribute greatly to uncertainties on 
the horizontally projected positions).  Uncertainties on the GPS positions are approximately 3 m (the measured lengths provide tighter constraints).  
Between 2003 and 2007, the area was logged, and a small logging road had been cut along the locations of the east, west, and central cameras.
}
\label{fig:camerapositions}
\end{center}
\end{figure*}

Knowing the precise relative positions of the cameras is necessary for extrapolating the camera energy measurements into a measurement of the 
total laser pulse energy at the ground.  During each observation, the location of each camera was measured using GPS, as well as a surveyor's tape measure to 
determine the relative positions via triangulation.  An example set of locations of the cameras, and their uncertainties, is given in Fig.~\ref{fig:camerapositions}.

The camera positions are overconstrained by the combination of GPS and measured distance information.  The measured distances in general provide tighter constraints than 
GPS, however the GPS measurements are needed to set the centroid position, and overall rotation angle, of the network.  To find the best-fit positions and their 
uncertainties, a $\chi^2$ minimization fit is performed using the ROOT graphical analysis program~\citep{ROOT}.

\subsection{CALIPSO Laser Pulse Energies}

CALIPSO measures the energy of each of its individual laser pulses.  The pulse energy monitoring on CALIPSO consists of NIST-calibrated 
photodiodes mounted on an integrating sphere, with 
pulse-by-pulse energy measurement with design absolute precision of $\pm2\%$ over full orbit, and relative precision of 
better than $\pm0.4\%$~\citep{win09}.  The pulse energies are recorded in the CALIPSO datasets available from the 
Atmospheric
Science Data Center (ASDC, located at NASA LaRC)\footnote{http://eosweb.larc.nasa.gov}, which may be compared with the pulse energies measured at the ground to obtain
measurements of atmospheric extinction.  The CALIPSO satellite-measured pulse energies which correspond to each of the six ground station observations
are given in Table~\ref{tab:observations}.

\begin{deluxetable}{lllccccc}
\tablecaption{\label{tab:observations} 
List of portable ground station observations of the CALIPSO beam in 2007 (with their geographic
positions and temperatures at the times of observation), the respective
measured laser pulse energies from the ground camera network ($E_{\rm ground}$) and from the CALIPSO satellite itself ($E_{\rm CALIPSO}$), as
well as the measured atmospheric extinction ratio ($r_e \equiv E_{\rm ground}/E_{\rm CALIPSO}$).}
\tablehead{
\colhead{Date} & \colhead{Lat.$^\ddagger$} & \colhead{Long.$^\ddagger$} & \colhead{$T_{\rm obs}$ ($^\circ$C)$^{\ddagger\ddagger}$} & \colhead{$E_{\rm ground}$ (mJ)} & 
\colhead{$E_{\rm CALIPSO}$ (mJ)} & \colhead{$r_e$} & \colhead{Notes} }
\startdata
13 Mar. & 49.090$^\circ$ N & 123.917$^\circ$ W & \phantom{0}5 $\pm$ 1 & ---                        & 111.3 $\pm$ 2.2 & ---                 & (1) \\ 
17 Apr. & 33.811$^\circ$ N & 111.996$^\circ$ W &           10 $\pm$ 3 & \phantom{0}41.1 $\pm$ 24.1 & 111.2 $\pm$ 2.2 & 0.4 $\pm$ 0.2       & (2) \\
28 Apr. & 37.606$^\circ$ N & 106.249$^\circ$ W &           $-1 \pm$ 3 & ---                        & 111.1 $\pm$ 2.2 & ---                 & (3) \\
1 May   & 41.012$^\circ$ N & 112.917$^\circ$ W &           10 $\pm$ 5 & 116.9 $\pm$ 52.4           & 111.2 $\pm$ 2.2 & $1.1^{+0.4}_{-0.5}$ & (4) \\
30 May  & 50.224$^\circ$ N & 126.561$^\circ$ W & \phantom{0}5 $\pm$ 3 & \phantom{0}77.7 $\pm$ 37.4 & 111.3 $\pm$ 2.2 & 0.7 $\pm$ 0.3       & (5) \\
1 Jun.  & 49.123$^\circ$ N & 123.931$^\circ$ W &           11 $\pm$ 3 & 108.8 $\pm$ 46.1           & 111.4 $\pm$ 2.2 & 1.0 $\pm$ 0.4       & (6) \\
\enddata

\vspace{0mm}
\begin{quote}
\noindent\underline{Notes:}

\hspace{1.6mm}$^\ddagger$\phantom{)} Location of the central camera during the observation.

\vspace{-1mm}$^{\ddagger\ddagger}$\phantom{)} Temperature data is from \citep{CanTempData} for the Canadian observations and from \citep{USTempData}
for the U.S.~observations.  Uncertainties are author-estimated due to the fact that the weather stations are not located in the same places
as the observations (\textit{e.g.}~the 1 May observation was nearly 200 miles from the nearest one), nor do all the closest stations have data at 
the times of each observation.

\vspace{-1mm}(1) Near Nanaimo, B.C., Canada.  Seen only in 1 camera (thus no ground energy measurement).

\vspace{-1mm}(2) Near Carefree, AZ, USA.  Full 7 camera observation.

\vspace{-1mm}(3) Near Monte Vista, CO, USA.  Icing on camera lenses (thus no ground energy measurement).

\vspace{-1mm}(4) Near the Great Salt Lake, UT, USA.  Full 7 camera observation.

\vspace{-1mm}(5) Near Woss, B.C., Canada.  Full 7 camera observation.

\vspace{-1mm}(6) Near Nanaimo, B.C., Canada.  Full 7 camera observation.
\end{quote}
\vspace*{-5mm}
\end{deluxetable}
\nopagebreak
\begin{deluxetable}{lcccc}
\tablecaption{\label{tab:systematics}Systematic uncertainties (in \% and in ${\rm mJ}$) on the measured pulse energies.}
\tablehead{
\multicolumn{1}{l}{Systematic Uncertainty} & 
\multicolumn{1}{c}{17 Apr.} & \multicolumn{1}{c}{1 May} & \multicolumn{1}{c}{30 May} & \multicolumn{1}{c}{1 Jun.} \\
\multicolumn{1}{l}{} & 
\multicolumn{1}{c}{(\% / mJ)} & \multicolumn{1}{c}{(\% / mJ)} & \multicolumn{1}{c}{(\% / mJ)} & \multicolumn{1}{c}{(\% / mJ)}
}
\startdata
Atmospheric Scintillation$^{\rm (a)}$                  & 46.7 / 19.2                     &   27.5 / 32.2                    & 32.7 / 25.4                     & 23.3 / 25.4    \\
CALIPSO Beam Profile Uncertainty$^{\rm (b)}$           & 35.0 / 14.4                     &   35.0 / 40.9                    & 35.0 / 27.2                     & 35.0 / 38.1    \\
Camera Throughput \& Absolute Calibration$^{\rm (c)}$  & 5.0 / 2.1                       &    5.0 / 5.8                     & 5.0 / 2.9                       & 5.0 / 5.4 \\
Camera/ADU Statistics$^{\rm (a)}$                      & 1.7 / 0.7                       &    0.8 / 0.9                     & 2.4 / 1.9                       & 0.6 / 0.7 \\
\noalign{\smallskip}
\hline
\noalign{\smallskip}
\textbf{Total}                                         & \textbf{58.6} / \textbf{24.1}  & \textbf{44.8} / \textbf{52.4}   & \textbf{48.2} / \textbf{37.4}  & \textbf{42.3} / \textbf{46.1}  \\
\enddata

\vspace{0mm}
\begin{quote}
\underline{Notes:}

As discussed in Section 6, the above uncertainties can be drastically reduced (to approximately 1.1\%) by (a) the addition of more cameras, with larger aperture,
(b) pre-flight precise measurement of beam profile and onboard monitoring,
and (c) improvements in the laboratory calibration of the camera optics.
\end{quote}
\vspace*{-2mm}
\end{deluxetable}


\begin{figure*}[p]
\begin{center}
\vspace*{-1cm}
\includegraphics[angle=0,width=8cm]{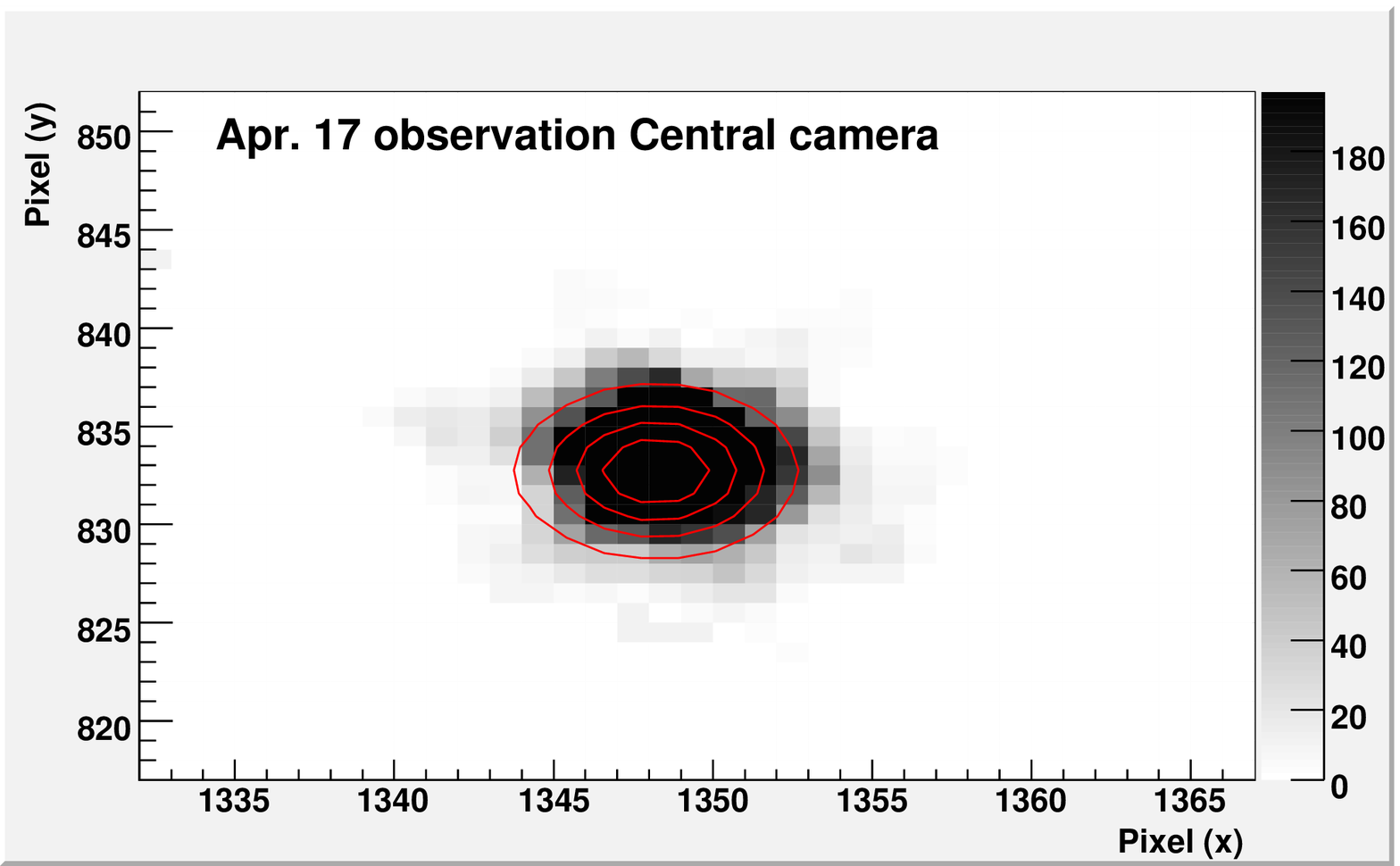}
\includegraphics[angle=0,width=8cm]{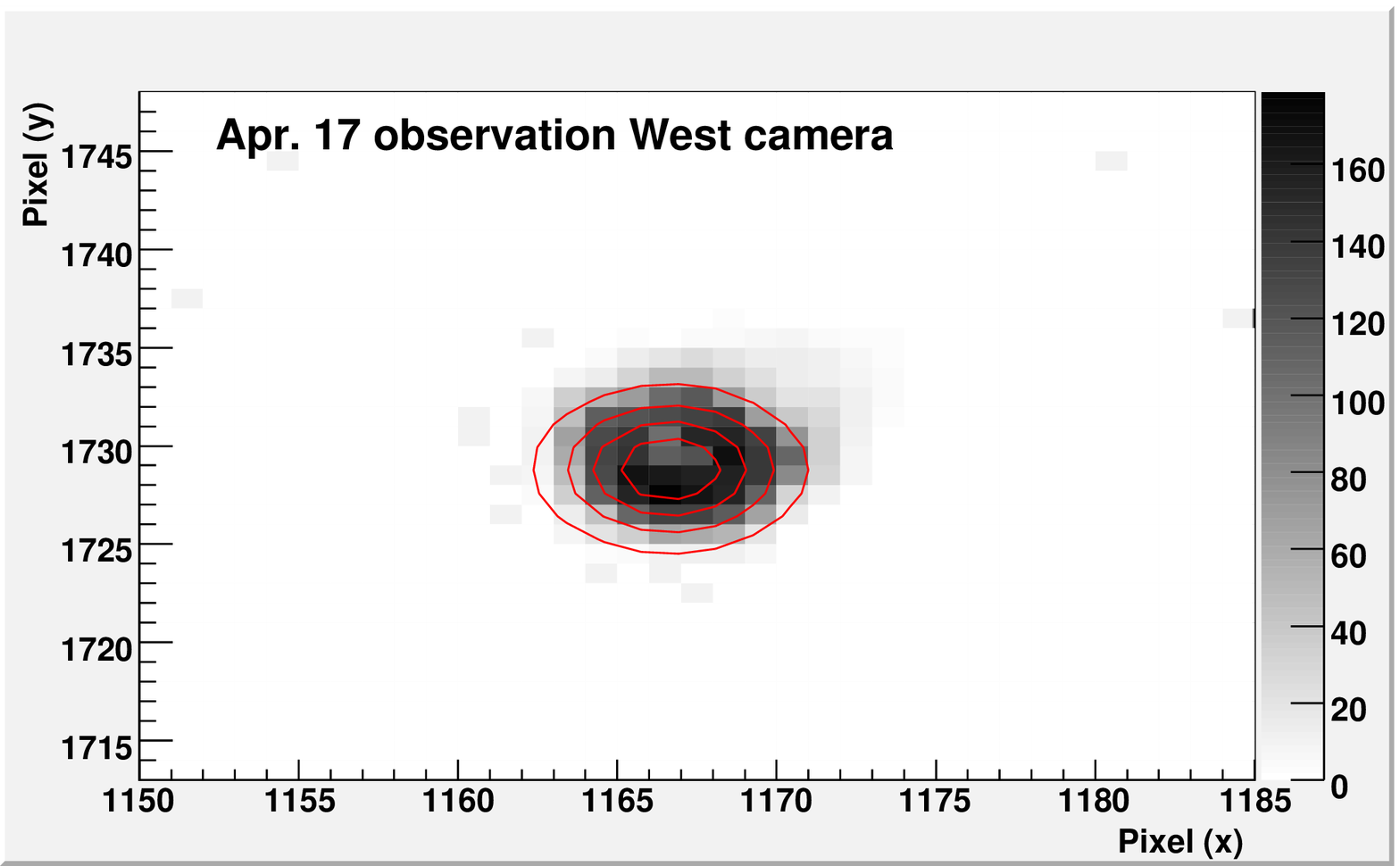}

\includegraphics[angle=0,width=8cm]{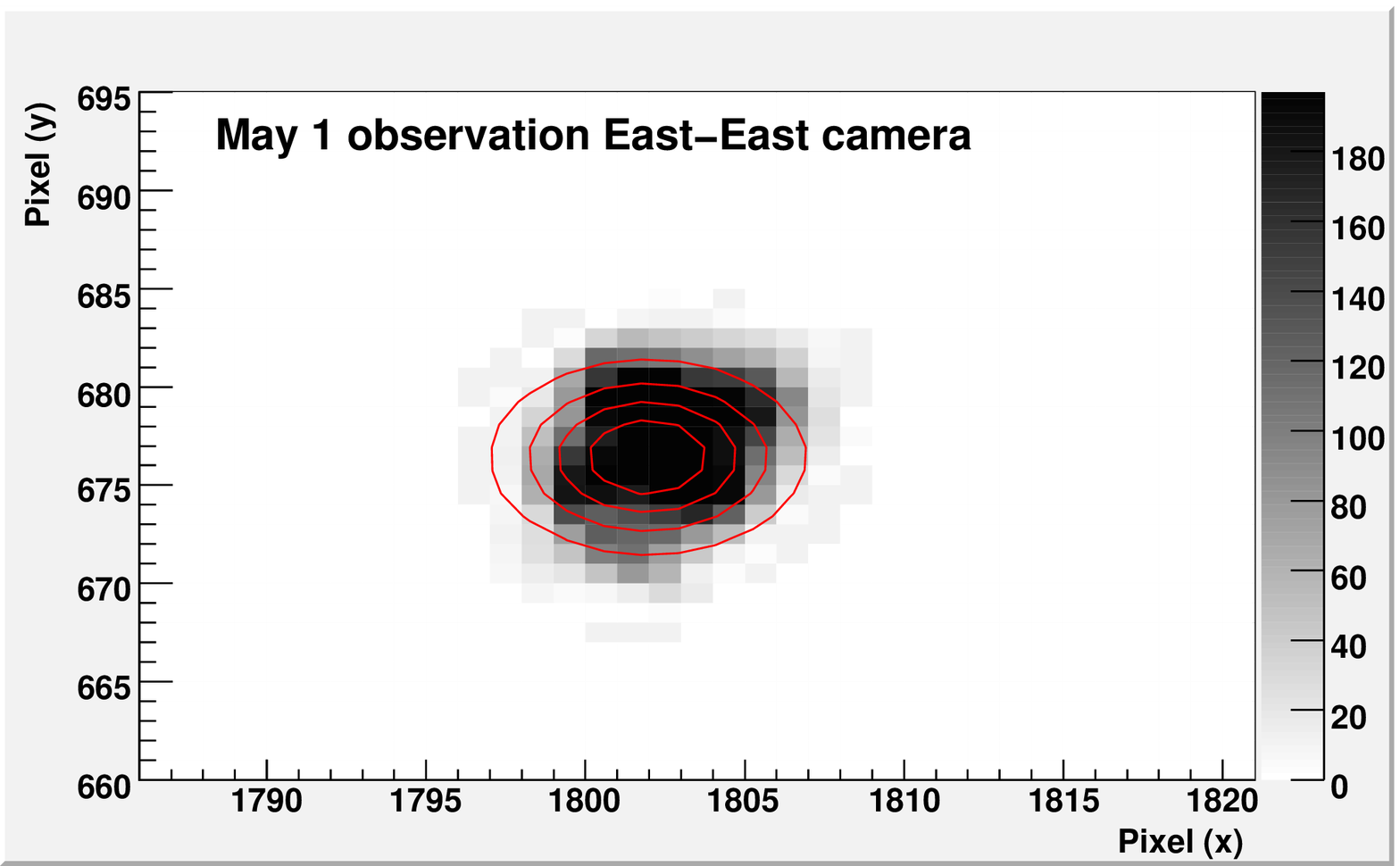}
\includegraphics[angle=0,width=8cm]{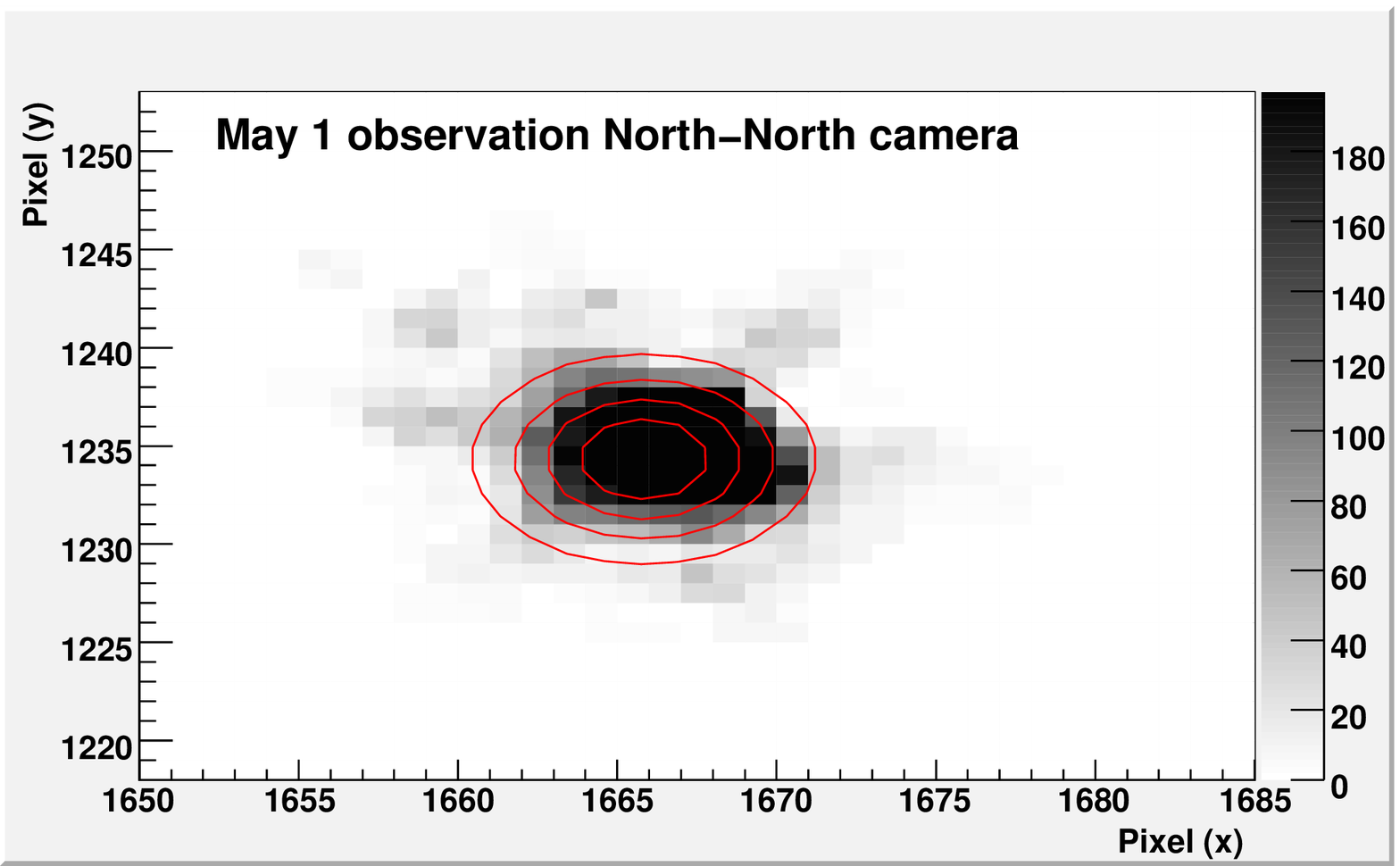}

\includegraphics[angle=0,width=8cm]{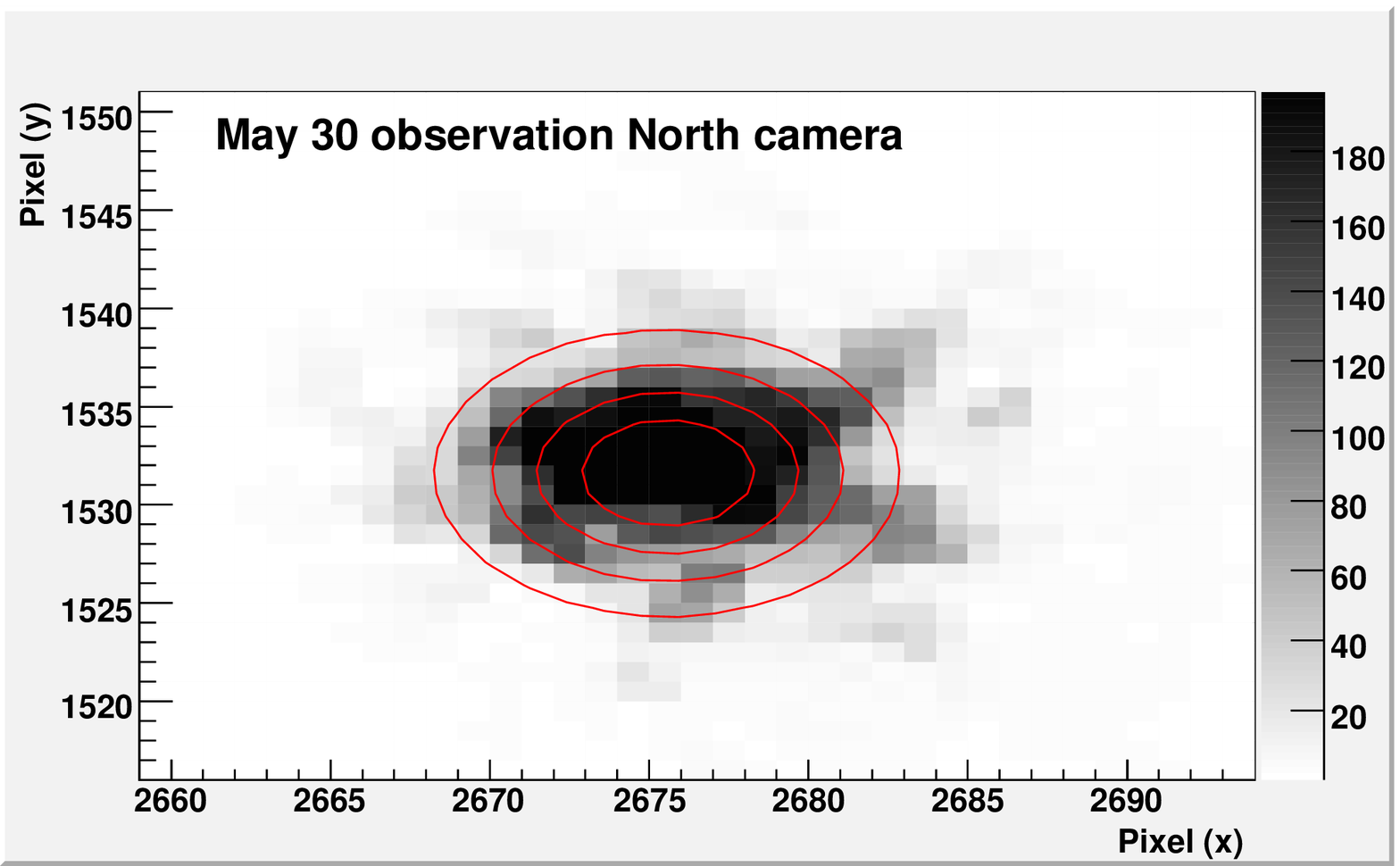}
\includegraphics[angle=0,width=8cm]{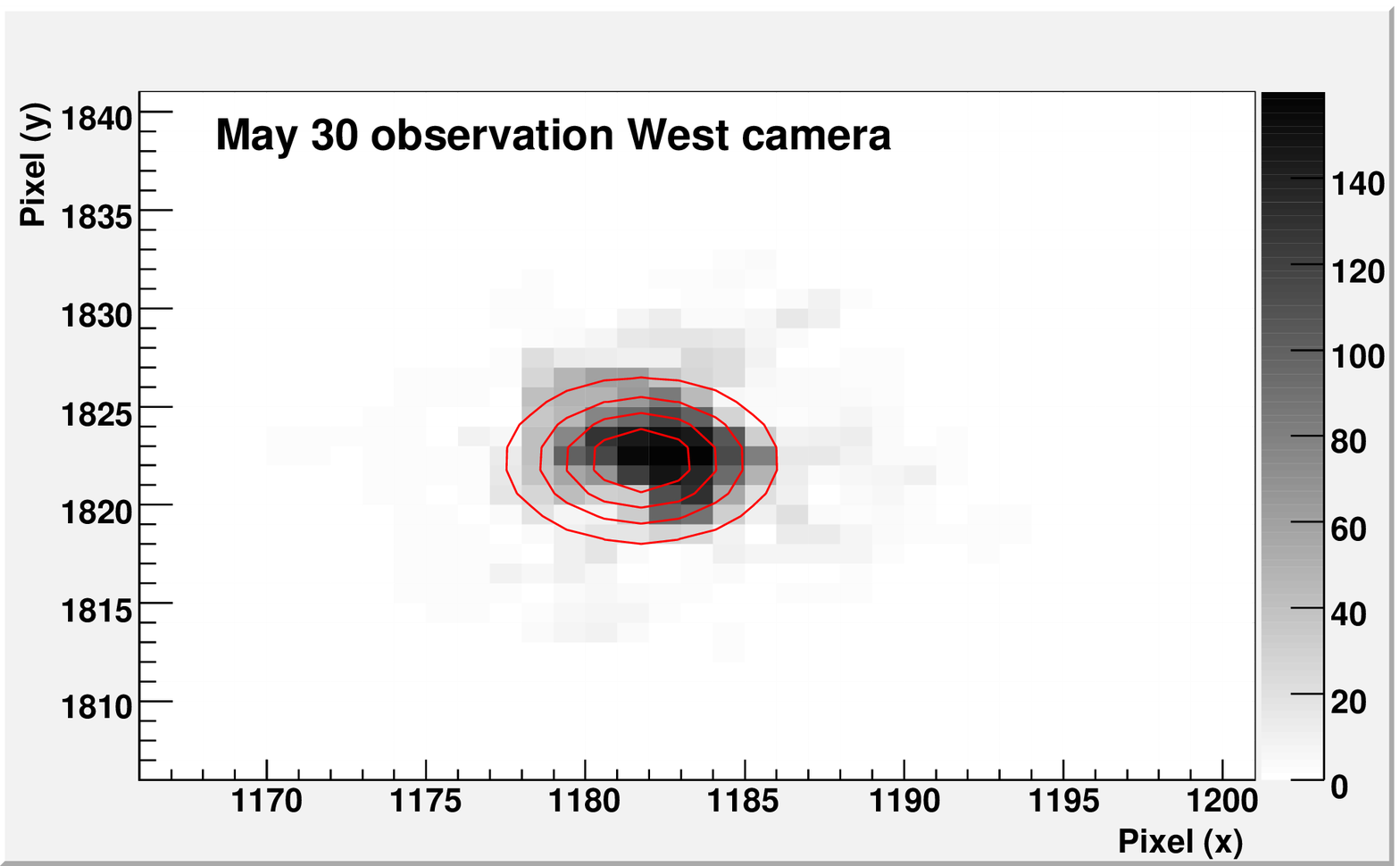}

\includegraphics[angle=0,width=8cm]{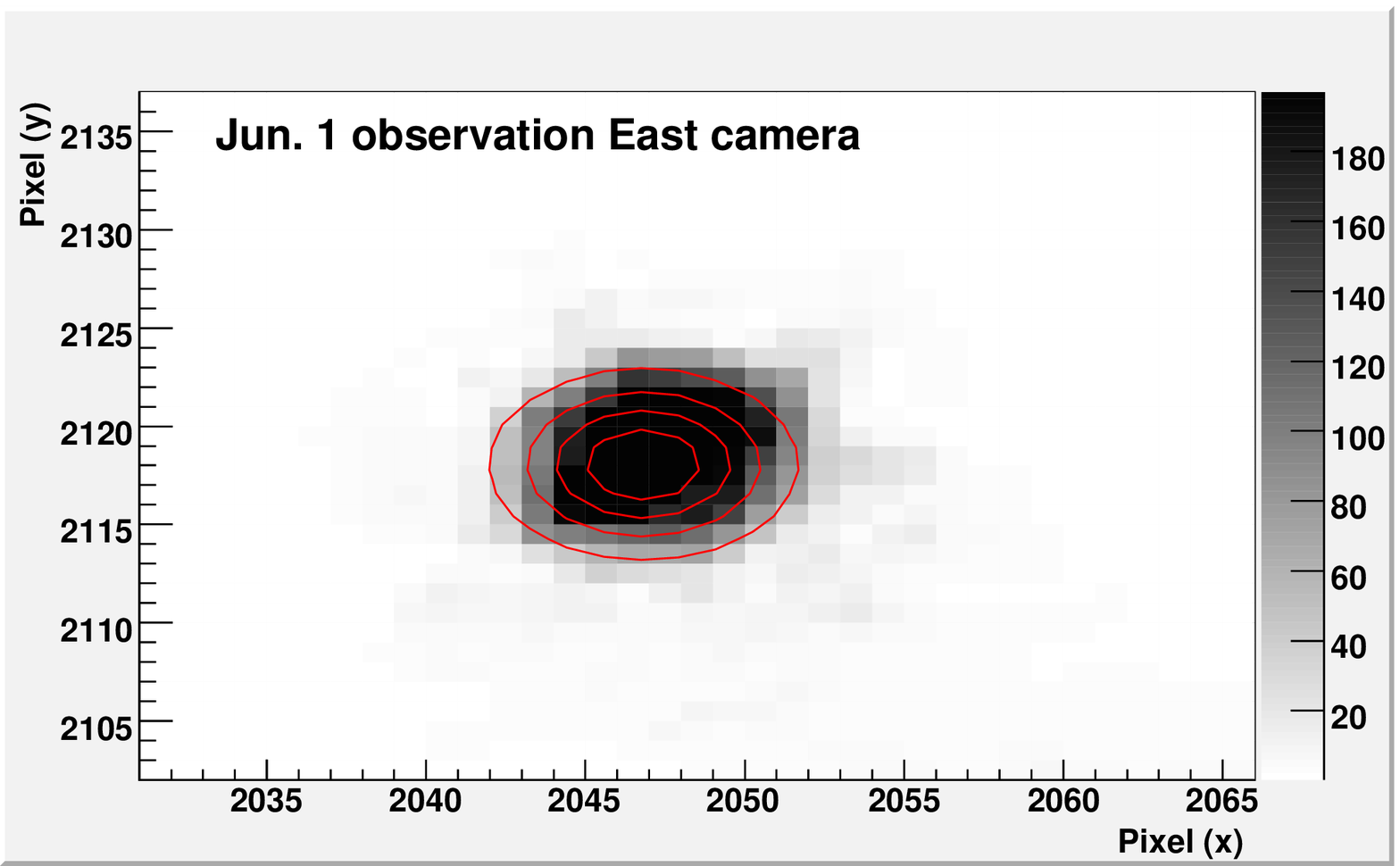}
\includegraphics[angle=0,width=8cm]{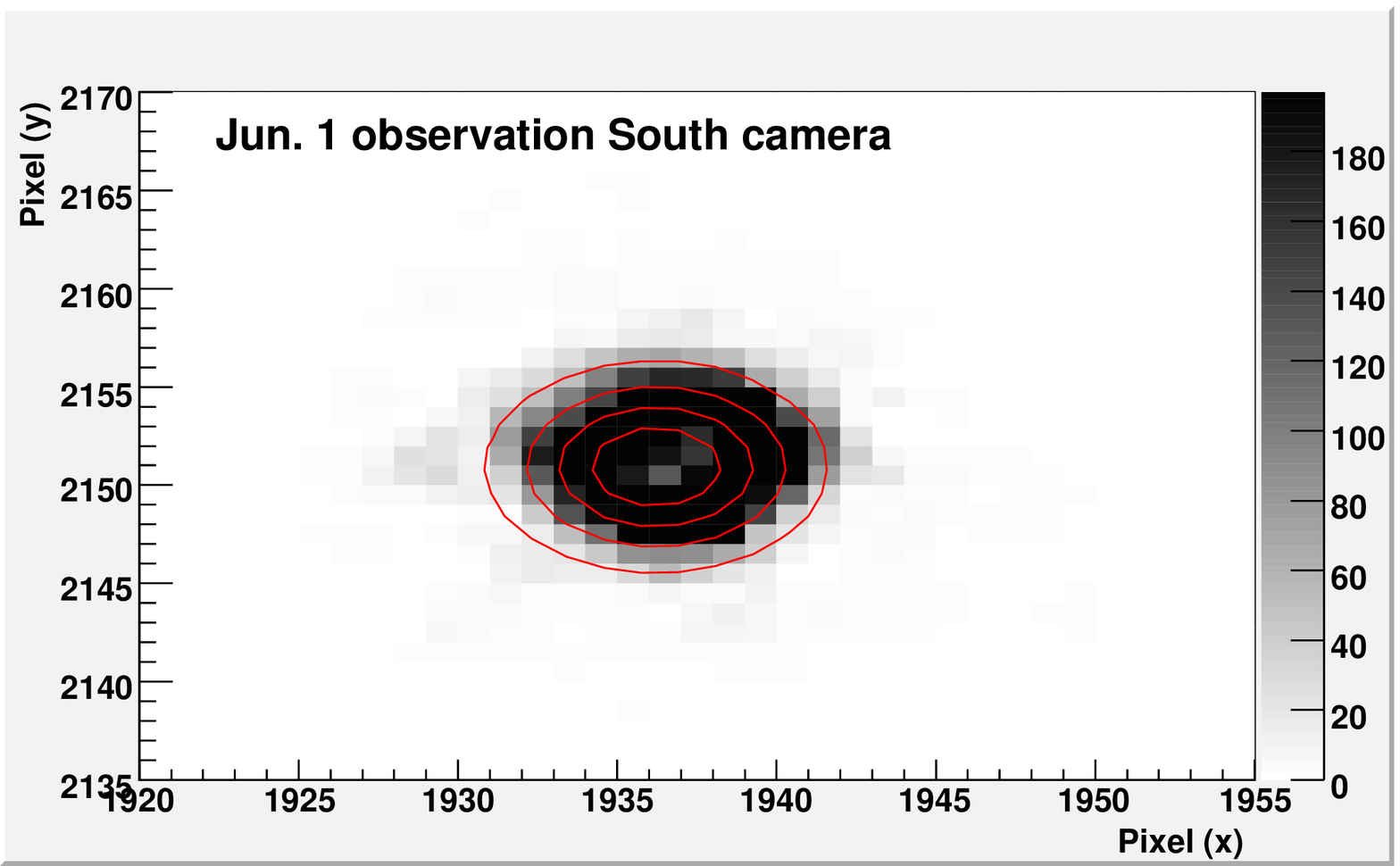}
\caption{
Camera image data for pixels in a 35 x 35 pixel region around recorded CALIPSO laser spots for eight example images.
Pixel intensities are in camera analog-to-digital units (ADUs).  The contours show the fitted 2-D Gaussian function. 
Pixels with values equal to the pixel saturation level (196) are considered in the fit to have true value either greater
than or equal to this value (such pixels have $\chi^2$ not constrained from above), as described in the text, thus the peak of
the fitted function can often exceed the saturation value.  There are 5 contours 
for each plot, equally spaced between the fitted function's maximum and minimum (usually 0) value in each histogram.
}
\label{fig:spotdata}
\end{center}
\end{figure*}

\begin{figure*}[p]
\begin{center}
\vspace*{-0.6cm}
\includegraphics[angle=0,width=15cm]{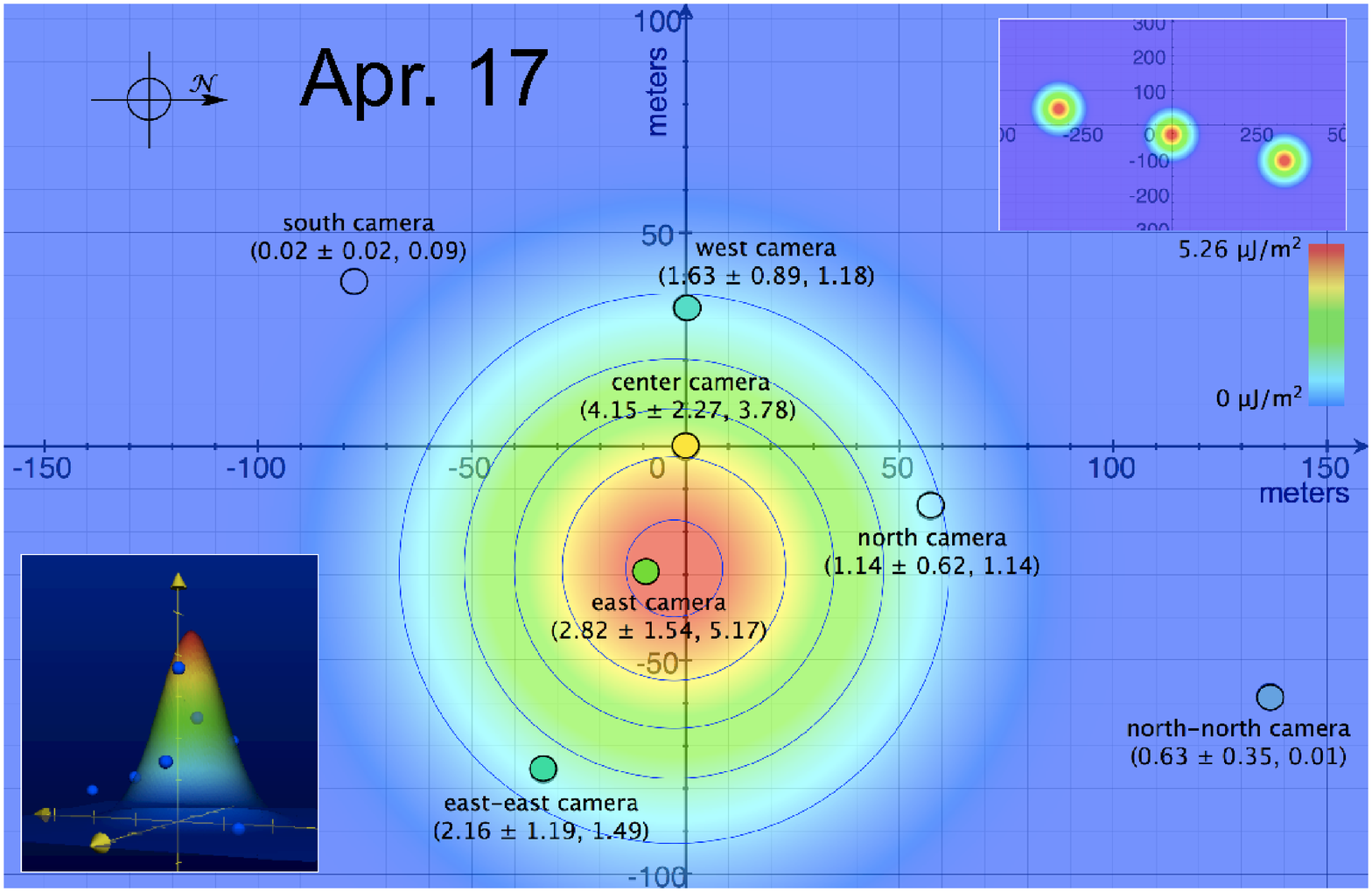}

\vspace*{1mm}

\includegraphics[angle=0,width=15cm]{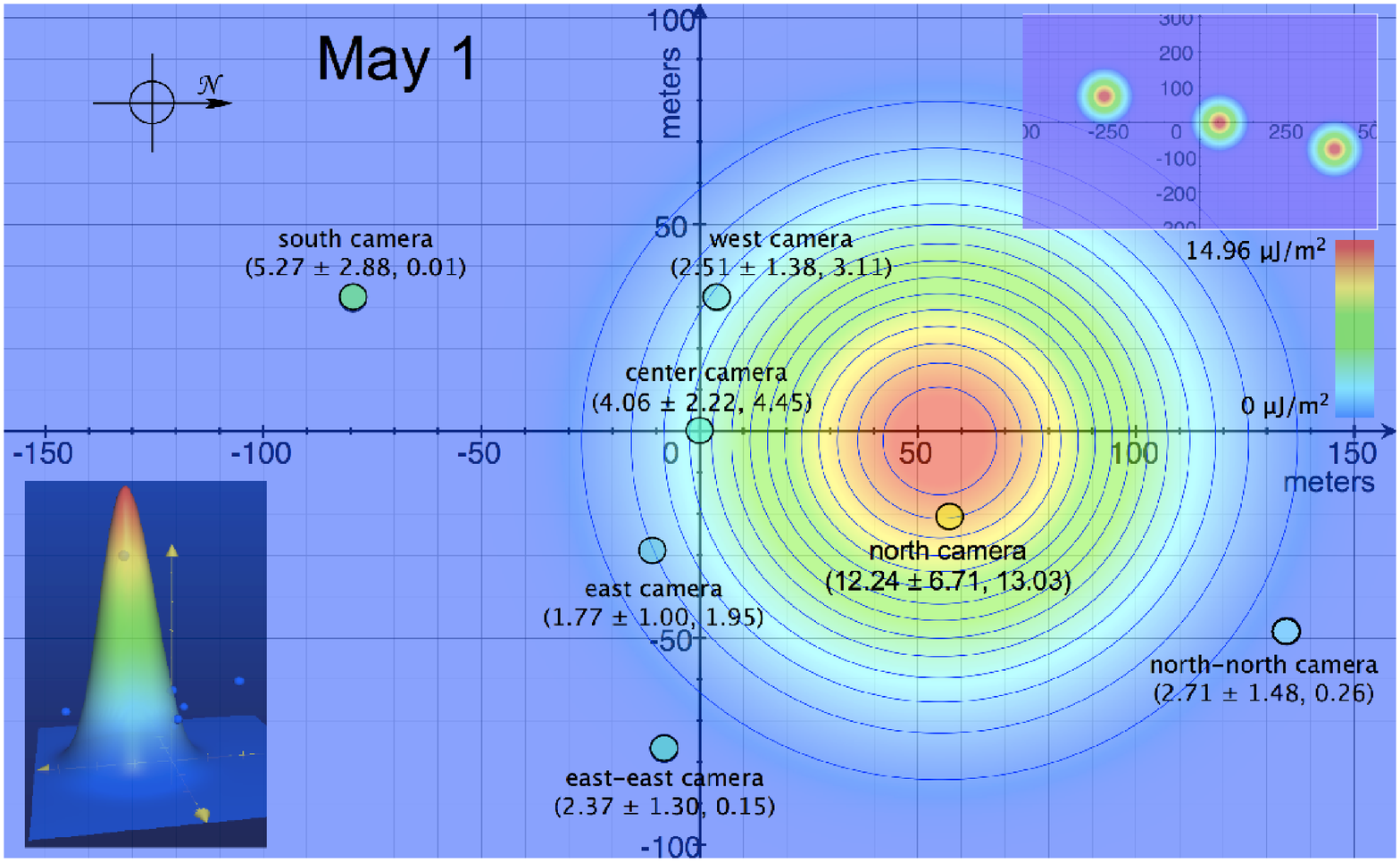}
\caption{
Camera time-integrated irradiance data, and the resulting fitted time-integrated irradiance maps, for the Apr. 17 (top) and May 1 (bottom)
observations.  The numbers at each camera refer to the measured value of time-integrated irradiance by that camera, with its associated
uncertainty (68\% CL), and the expectation value from the fitted function at the location of that camera.
The contours on each plot are spaced at 1 $\mu$J/m$^2$ intervals.
The upper-right inset on each plot extends the $x$ and $y$ axes in order to see the three 2-D Gaussians that comprise the 
fitted function (as described in the text), and the lower-left inset shows a different 3-D view (from the side, rather than from above) 
of the fitted function and the data points.
}
\label{fig:irradmaps1}
\end{center}
\end{figure*}

\begin{figure*}[p]
\begin{center}
\vspace*{-0.6cm}
\includegraphics[angle=0,width=15cm]{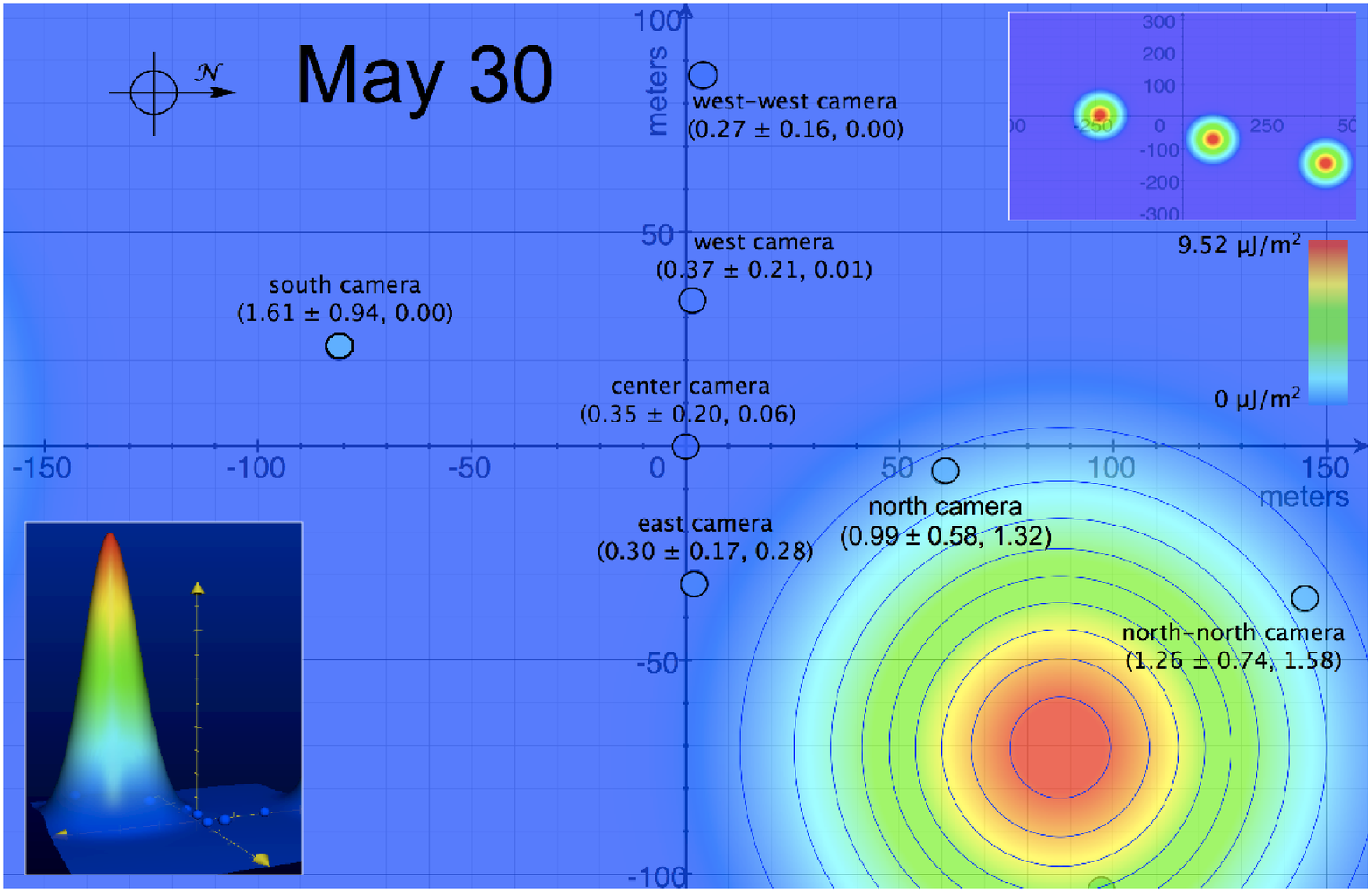}

\vspace*{1mm}

\includegraphics[angle=0,width=15cm]{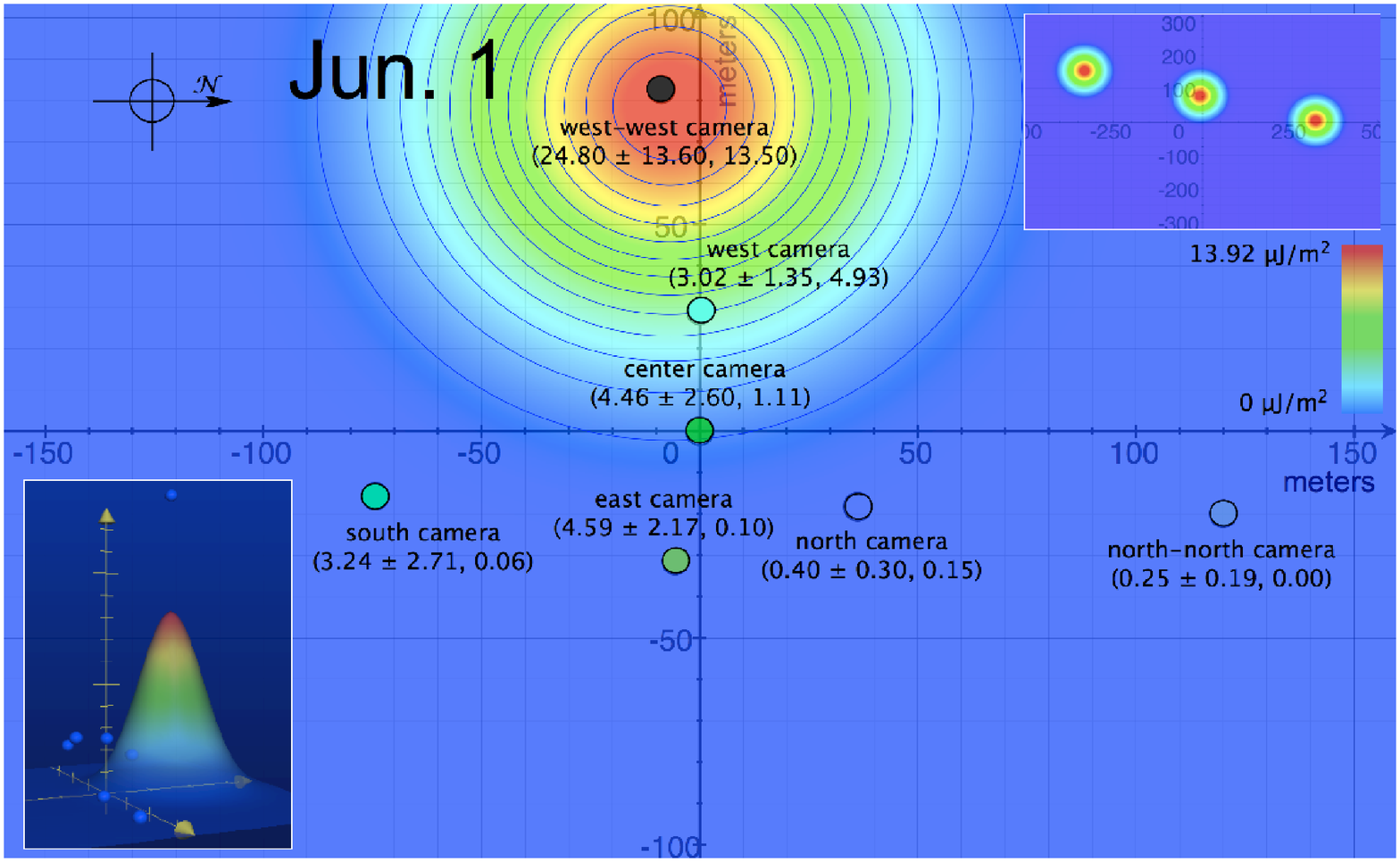}
\caption{
Camera time-integrated irradiance data, and the resulting fitted time-integrated irradiance maps, for the May 30 (top) and Jun.~1 (bottom)
observations.  Please see the caption of the figure on the previous page for a description of these plots.
In both of these two observations, the location (foliage, etc.)~prevented a second camera from being placed further east than the 
east camera, thus a ``west-west'' camera location was used instead of an ``east-east'' one.  (Also note that central value of the measured 
time-integrated irradiance at the west-west camera location on the Jun.~1 observation is above the color scale chosen for that plot.)
}
\label{fig:irradmaps2}
\end{center}
\end{figure*}

\subsection{Extinction Results}
\label{results}

The image data text file produced for each observation from each camera, as described above, must be fitted to a function that describes the CALIPSO laser spot
as well as background from scattered light etc.~(only rarely did any stars happen to come within the 51 pixel square, corresponding to about
0.1 square degrees, that contained the CALIPSO spot).  The data were found to be consistent with a single 2-D Gaussian describing the spot, plus a uniform background.
However, for images in which the spot is especially bright, the center pixels often reach the saturation limit of the camera.  Thus, to fit the data, a likelihood
function that allows for pixel saturation (merely constraining the fitted function to be anything \textit{above} the saturated pixel values, rather than the values
themselves, in the case of the pixels that have reached the saturation limit --- and otherwise being a standard $\chi^2$ likelihood) must be used.  When combined with the 
Gaussian plus uniform background, this likelihood function describes the data well.

For each image data text file, a total of five variables are floated in the fit, which is performed in ROOT~\citep{ROOT}: the $x$ and $y$ positions of the spot centroid, the 
standard deviation of the Gaussian spot, the level of uniform background, and the total amount of light in the CALIPSO spot.  Asymmetric Gaussian uncertainties are 
determined for all variables.

Plots of pixel data in the region surrounding the CALIPSO spot in eight of the camera images can be found in Fig.~\ref{fig:spotdata}.  Further plots and data may be found
online at .

The above fits provide a measurement of the amount of light recorded in each camera, in the digital pixel units of the camera.  To translate these digital unit measurements 
into SI units of energy, we use the absolute photometric calibration data for each camera as described above.  This provides us with absolute energy measurements for each
camera (with additional uncertainty due to the calibration-related error budget).

The above procedure determines the amount of light reaching each individual camera.  To determine the total amount of CALIPSO light reaching Earth in a given observation, 
the data from all seven cameras must be combined together (along with the fitted positions of the cameras as described above).  Thus, a secondary fit, again using ROOT, is 
performed which combines this information into a measurement of the total amount of light from a CALIPSO pulse reaching Earth.  In this fit, the fact that more than one 
pulse may contribute to the light measured in the seven cameras must be taken into account.  A function consisting of three Gaussian pulses is fitted to the data from the 
seven cameras.  (Note that a train of three pulses extends far further than the size of the camera network.)  The direction heading of the CALIPSO satellite, and the 
separation between CALIPSO pulses, are known well from NASA ground track information and the
precise 20.25 Hz frequency of the laser respectively.  However, the centroid on Earth of a given pulse (in both latitude and longitude), as well as its total amount of 
light, must be fitted.  Thus, those three variables are floated in the fit, and central values and asymmetric uncertainties are determined for each.  The results of these
fits are shown in Figs.~\ref{fig:irradmaps1} and~\ref{fig:irradmaps2}.

The measured energies of the CALIPSO pulses reaching Earth, denoted as $E_{\rm ground}$, are given in Table~\ref{tab:observations}.  The
CALIPSO satellite-measured pulse energies, $E_{\rm CALIPSO}$, are also given, as well as their ratios $r_e$ which provide measurements of extinction.  

Although the uncertainties are large, the results, with the possible exception of the Apr.~17 observation, are consistent with the expectation of $r_e \approx$ 0.95--1.0 (given 
the fact that all observations were made on relatively clear nights).
The low value for the Apr.~17 observation may be due to the fact that this observation was taken in a desert environment, and blowing dust might 
have partially obscured the camera lenses---especially since the cameras were not placed on tripods for that initial observation.

The breakdown of the respective uncertainties
on the measurements of energy reaching the ground are given in Table~\ref{tab:systematics}.  As one can see, atmospheric scintillation, calculated as per~\citet{fri67}, and 
uncertainty in the profile shape of the CALIPSO beam [the beam is assumed to be a 2-D Gaussian in shape; the uncertainty on the true shape is taken to represent a $\pm$35\% uncertainty
on the measured energy value~\citep{win09}] are by far the dominant uncertainties.
The former can be
reduced by increasing the aperture and the number of cameras, and 
precise pre-flight measurement of beam profile (along with onboard monitoring)
can dramatically reduce the latter uncertainty.  Increasing the number of cameras (and/or the bit depth of their CCDs) can also reduce the small uncertainty due to ADU 
statistics.  The other relatively small uncertainty due to camera throughput and absolute calibration (taken to be $\pm$5\% of the measured energy value) can be reduced by functionality 
and use of calibrated photodiode measurements during observations, as well as with more thorough laboratory studies of the cameras.

\begin{figure*}[t]
\begin{center}
\includegraphics[angle=0,width=18cm]{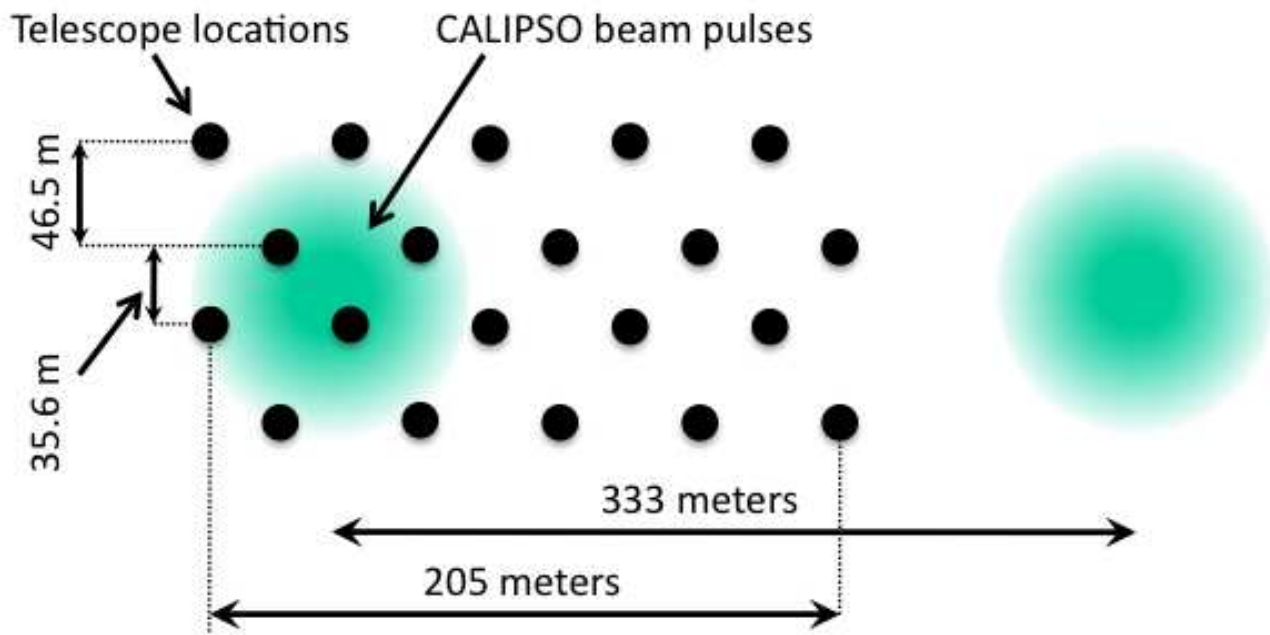}
\caption{
Optimized locations for the network of twenty 16'' telescopes.  The CALIPSO pulse locations are, of course, only expectation values:
the location of the pulses has a Gaussian uncertainty of 50 m transverse to the satellite direction of motion, and the phase of the 
pulses in the train (the longitudinal location) is entirely uncertain (only the distance between the pulses is known).  The optimum
length of the telescope network (as shown in the optimization plots in Fig.~\ref{fig:GridLenOptim}) is less than the separation
distance between CALIPSO pulses due to the importance of having as many telescopes as possible observing a pulse (despite the 
relatively small risk of having the entire network fall in the gap between pulses in a given observation).
}
\label{fig:hexnetwork}
\end{center}
\end{figure*}

One could compare these results with independent extinction measurements based on the measured light from stellar standard sources within the camera images during 
observation, to determine their consistency.  However, in light of the very large atmospheric scintillation and 
beam profile-related
uncertainties on the CALIPSO-based measurements, such
a comparison would be more fruitful with a future more precise and larger telescope observation network, and the next-generation LIDAR satellite following CALIPSO.


\section{MONTE CARLO STUDIES OF OPTIMAL TELESCOPE NETWORK DESIGN}
\label{MCstudies}

In order to reduce the dominant uncertainty due to atmospheric scintillation, significantly larger apertures than provided by the small 
Panasonic cameras are required for aperture averaging over atmospheric cells.  In 2009-10, a study was done to optimize the design of an 
entirely new network of telescopes and cameras to provide the best possible measurements of CALIPSO laser light energy reaching the Earth, 
at a given equipment cost.  A Monte Carlo simulation program to perform this optimization was developed. 

Consumer 16'' aperture Dobsonian reflector telescopes are available at relatively low cost, 
and can provide an accessible means of 
aperture-averaging over atmospheric scintillation with a pulsed laser satellite source.  Compared with a relative standard deviation
of observed energy due to atmospheric scintillation of $\sigma_I = 0.466$ for small aperture cameras (such as the Panasonics) with pulsed 532 nm light
(as described in Sec.~2), the relative standard deviation for a 16'' aperture telescope is $\sigma_I = 0.035$
[calculated as per~\citet{fri67}], a reduction factor far outweiging the increase in cost.

We consider the optimization, for a given observation attempt of a CALIPSO pulse, of telescope locations in a network of 20 such telescopes.
The uncertainty on the predicted path of CALIPSO is approximately 50 meters in the direction orthogonal to the satellite's travel (which, for
the purposes of this and the next paragraph, we will denote by the $y$ direction, with the direction of the satellite's travel denoted as the $x$ direction).
We take that uncertainty to be Gaussian distributed.  CALIPSO pulses are separated by 333 m.  In the $x$ direction, the location
of the pulses is entirely uncertain \textit{a priori}; only their separation is known.\footnote{In principle, this may be able to be predicted to
some degree by studying the timing of the pulses from 3 days earlier (CALIPSO data takes 3 days to become available for analysis) and extrapolating
via the 20.25 Hz pulse rate, but it is likely that this would not provide precise information on the position of the pulses.}  Thus no advantage is gained by having an
elliptical distribution of the telescopes, centered around some point in the $x$ direction, rather than a rectangular distribution.
Thus, we consider a grid of telescopes, either a 4 x 5 rectangular pattern or 4 x 5 hexagonal pattern (\textit{e.g.}~such as that shown in 
Fig.~\ref{fig:hexnetwork}), with 4 telescopes in the $y$ direction.  (Other arrangements, such as 5 x 4 or 2 x 10, were also considered; these are
considerably less effective.)  In the $y$ direction, the optimal placement of the telescopes is such that they are evenly distributed in the space 
corresponding to the integral of the probability density distribution of the irradiance of the beam pulses.  The
probability density of the satellite path is a Gaussian distribution with $\sigma$ = 50 m, and the pulses themselves are Gaussian with $\sigma$ = 35.25 m,
thus the probability density distribution of the $y$ projection of the irradiance is also Gaussian with $\sigma$ equal to the addition in quadrature of
50 and $35.25/\sqrt{2}$, \textit{i.e.}~55.87 m (with the factor of $\sqrt{2}$ due to the projection of the 2-D Gaussian pulses onto the 1-D $y$ axis).
The optimal placement of the telescopes will be where the integral of that Gaussian reaches 1/8, 3/8, 5/8, and 7/8; thus creating an even spread over the
normal distribution.  This occurs when the telescopes are 17.8 m and 64.3 m to the right and left (in the $y$ direction) of the expectation value
of the satellite path, corresponding to 46.5 m and 35.6 m separation between the telescopes in $y$, as shown for example in Fig.~\ref{fig:hexnetwork}.
This sets the $y$ distribution.  However, the placement of the telescopes in $x$, and whether to use a rectangular or hexagonal grid, must still
be determined by simulation.

\begin{figure}[!t]
\begin{center}
\includegraphics[angle=0,width=7.7cm]{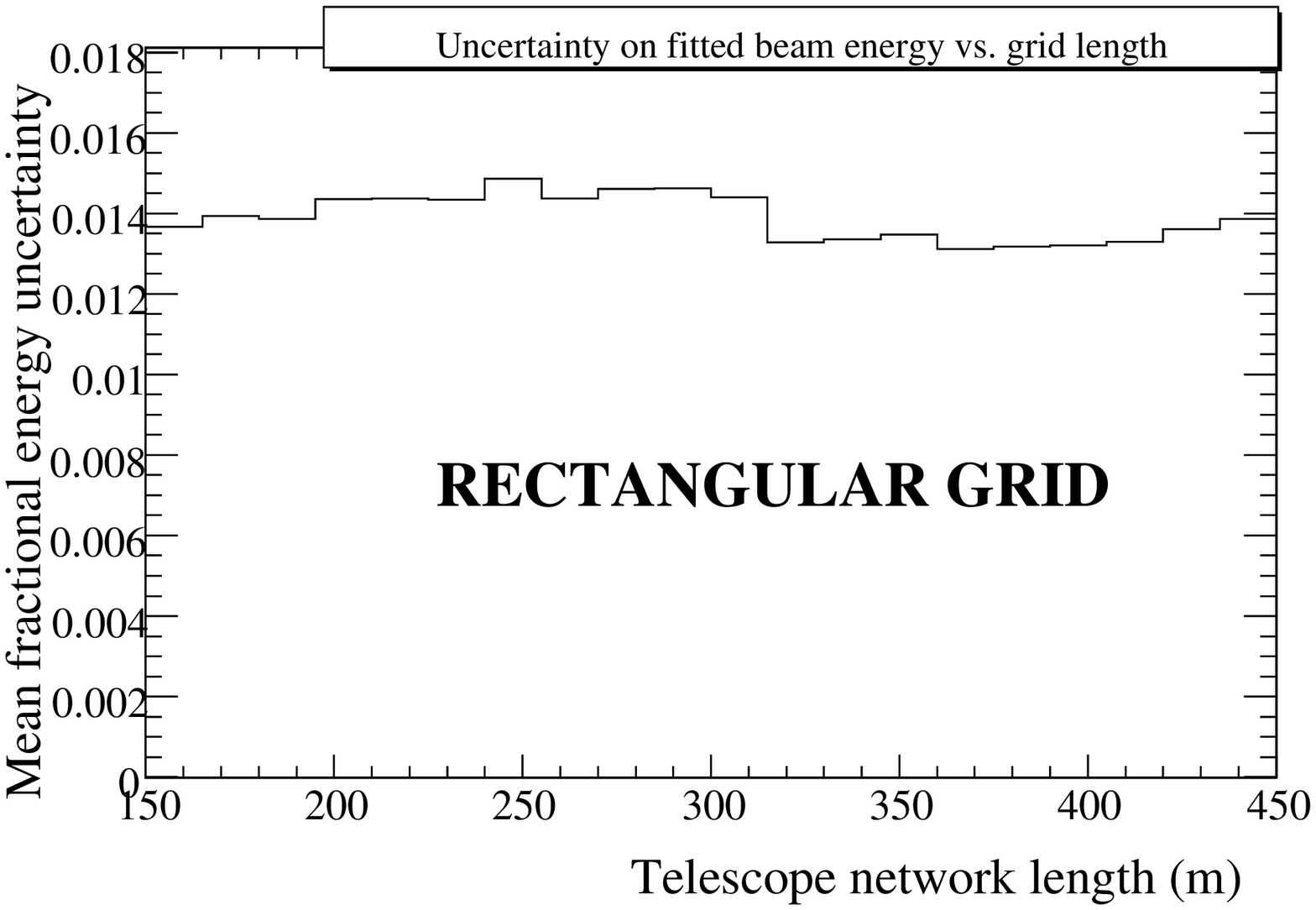}
\includegraphics[angle=0,width=7.7cm]{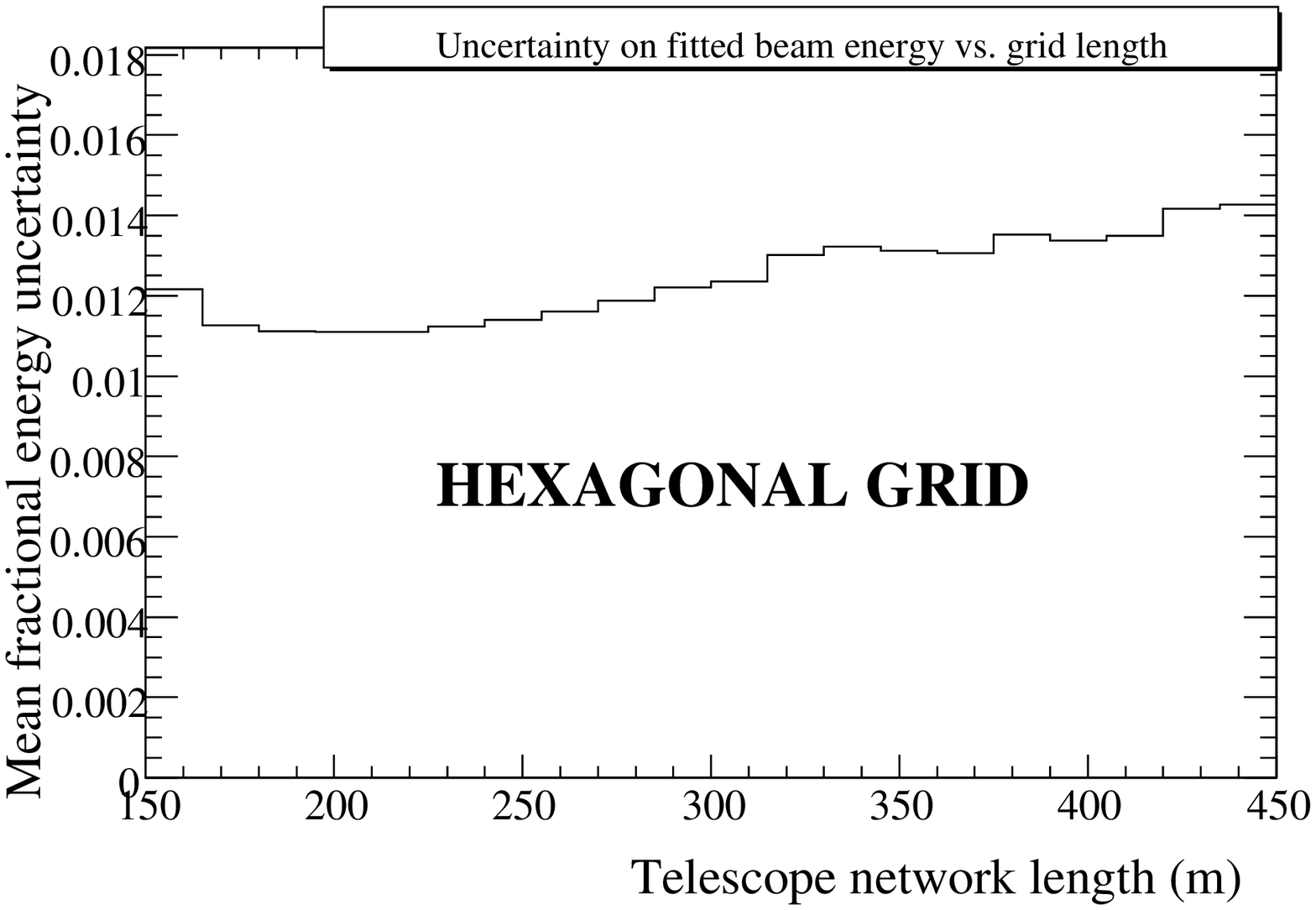}
\caption{
The average uncertainty, per observation, on measurements of the CALIPSO beam pulse energy reaching the ground (as a fraction
of that energy) for (top) a rectangular 4 x 5 grid of twenty 16'' telescopes and (bottom) a hexagonal grid of twenty 16'' telescopes
(as shown in Fig.~\ref{fig:hexnetwork}), as functions of the length of the grids in the direction of the CALIPSO satellite motion.
This uncertainty includes effects due to atmospheric scintillation and camera/ADU statistics, but does not include absolute
calibration or beam profile effects.  The values were determined using 5000 MC ``observations'' for each of the twenty bins included on
each plot.  As seen above, the optimum grid pattern for the 20 telescopes is a hexagonal grid 205 meters in length.
}
\label{fig:GridLenOptim}
\end{center}
\end{figure}

\begin{figure}[!t]
\begin{center}
\includegraphics[angle=270,width=7cm]{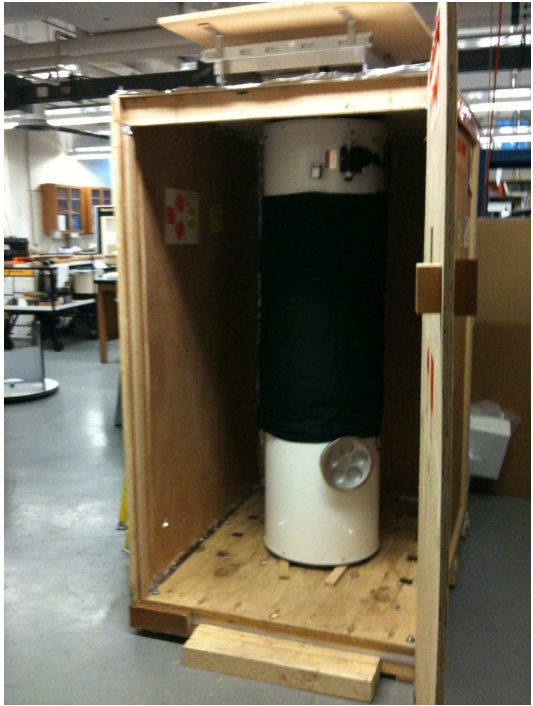}
\caption{
Calibration dark box for the 16'' telescope.  A 2-D moveable stage for an attenuated 532 nm laser beam for calibration is on the top of the box.
}
\label{fig:CalibBox}
\end{center}
\end{figure}

Using the ROOT software package~\citep{ROOT}, a Monte Carlo simulation program that incorporates the effects of atmospheric scintillation and camera/ADU
statistics, and allows for the variation of placement of telescopes in patterns around the satellite ground track, was developed.  Plots of the average
uncertainty, per observation, on the measured CALIPSO beam pulse energy reaching the ground, for rectangular and hexagonal 4 x 5 grids of 16'' telescopes,
as functions of the length of the telescope network in the $x$ direction, are shown in Fig.~\ref{fig:GridLenOptim}.  
It is seen that the minimum uncertainty per observation is obtained with a hexagonal grid of length 205 m, as shown in Fig.~\ref{fig:hexnetwork}.  As shown in the
plot, the average uncertainty per observation for this pattern is approximately 1.1\%.  This includes effects of atmospheric scintillation and camera/ADU
statistics, but does not include uncertainty due to laboratory and on-site absolute calibration of the telescopes and cameras, which are effects that at this
level of uncertainty would be likely to dominate in the absence of a very careful program for absolute calibration.  The simulation also includes the effect
of observations which fail to yield a reconstructible measurement of the observed energy for a CALIPSO pulse 
(approximately 4.5\% of observations with the optimized
setup), primarily due to the network falling in between (in $x$), or to the left or right (in $y$), the train of CALIPSO pulses; such failed observations
contribute zero to the average of the inverses of the standard deviations of the simulated observations.

In order to test the operation of a network of larger telescopes, a single Meade 16'' telescope, as shown in Fig.~\ref{fig:CalibBox}, was added to the center of the 
camera and photodiode network.  This telescope has been sucessfully operated, together with the network of the seven small Panasonic cameras,
at a single CALIPSO overpass near Comox, B.C., in July of 2009.



\section{FUTURE OF THE GROUND-BASED OBSERVATION NETWORK}
\label{FutureNetwork}

As shown above, the uncertainties on atmospheric extinction in the measurements using the Panasonic camera network are dominated by atmospheric scintillation, 
with the second-largest source being 
beam profile-related
uncertainty.  Both of these uncertainties can be dramatically reduced by using
larger-aperture devices (as described in the MC studies above), and by 
precise pre-flight measurement of beam profile shape.
However, moving a network of 16'' telescopes poses some 
logistical challenges that are not present with a network of small consumer cameras.  Storage, transport, and setup facilities are required 
with a number of larger devices.  Such challenges are certainly not insurmountable, though.  As an intermediate step, a smaller network of
six 16'' telescopes, together with the present Panasonic cameras, is more easily transportable, and would reach a precision of
6.0\% average uncertainty per observation (when the significantly increased level of observations that fail to yield a reconstructable measurement
is taken into account), a little over a factor of 5 less precise than a twenty telescope network.

We envision a small storage area for the telescopes and other hardware
in the southwestern U.S., with facilities for moving and setup of the telescope network below an expected satellite overpass, 
with each setup for a twenty 16'' telescope network as shown in Fig.~\ref{fig:hexnetwork}.
As LIDAR-based satellites for atmospheric science will continue to be available following the completion of the CALIPSO mission (see below), and
as the geo-location of laser pulses is important for such missions (in addition to the motivation of atmospheric calibration), it is 
envisioned that the time and manpower necessary for such a facility will become available.

\section{FUTURE OF SATELLITE-BASED CALIBRATION STANDARDS}
\label{FutureSatellites}

Following the completion of CALIPSO's mission, it is envisioned that, as per recommendations of the National Academy of Sciences,  
a follow-up mission, known as ACE (Aerosol-Cloud-Ecosystems), will succeed CALIPSO, and
continue to provide a well-calibrated source of visible light from low Earth orbit~\citep{NAS_ACE}.  ACE is forseen to have
an increase in pulse frequency above that of CALIPSO, to 100 Hz, which will improve the ability to measure pulses from the ground, as
the pulses will not be so widely separated.  Thus, the ground-based network can be made more compact, and/or one will have the ability
to measure multiple pulses per overpass, increasing the data sample.  With a network of twenty 16'' telescopes as shown above, 
we expect to obtain uncertainties on extinction, and thus absolute photometry, to better than 
1.1\% for multiple pulses from each observation of ACE, limited by only by the precision of laboratory and on-site absolute calibration of the telescopes and cameras.

However, such measurements would still be valid only for the zenith, at the specific location of the telescope network, for the
specific laser frequency.  To obtain results that apply more directly to calibration of photometry for the majority of
astronomical results, the satellite laser source should be able to
point at major observing facilities, and modify its frequency to be able to scan through the visible and near-infrared spectrum.  Both
capabilities are definitely within reach of technology, as lasers that are tunable from the near-IR, through the visible, through near-UV,
as well as movable mirrors to point a beam, are presently common features of laboratories and could be qualified for space.  However, such capabilities will require the 
generation of hardware beyond ACE.  Both pointability (to scan over temporal atmospheric features, \textit{e.g.}~hurricanes, forest fires, and man-made sources of 
pollution), and wavelength tunability, are of interest to CALIPSO and ACE atmospheric scientists as well.

\begin{figure}[!t]
\begin{center}
\includegraphics[angle=0,width=7.7cm]{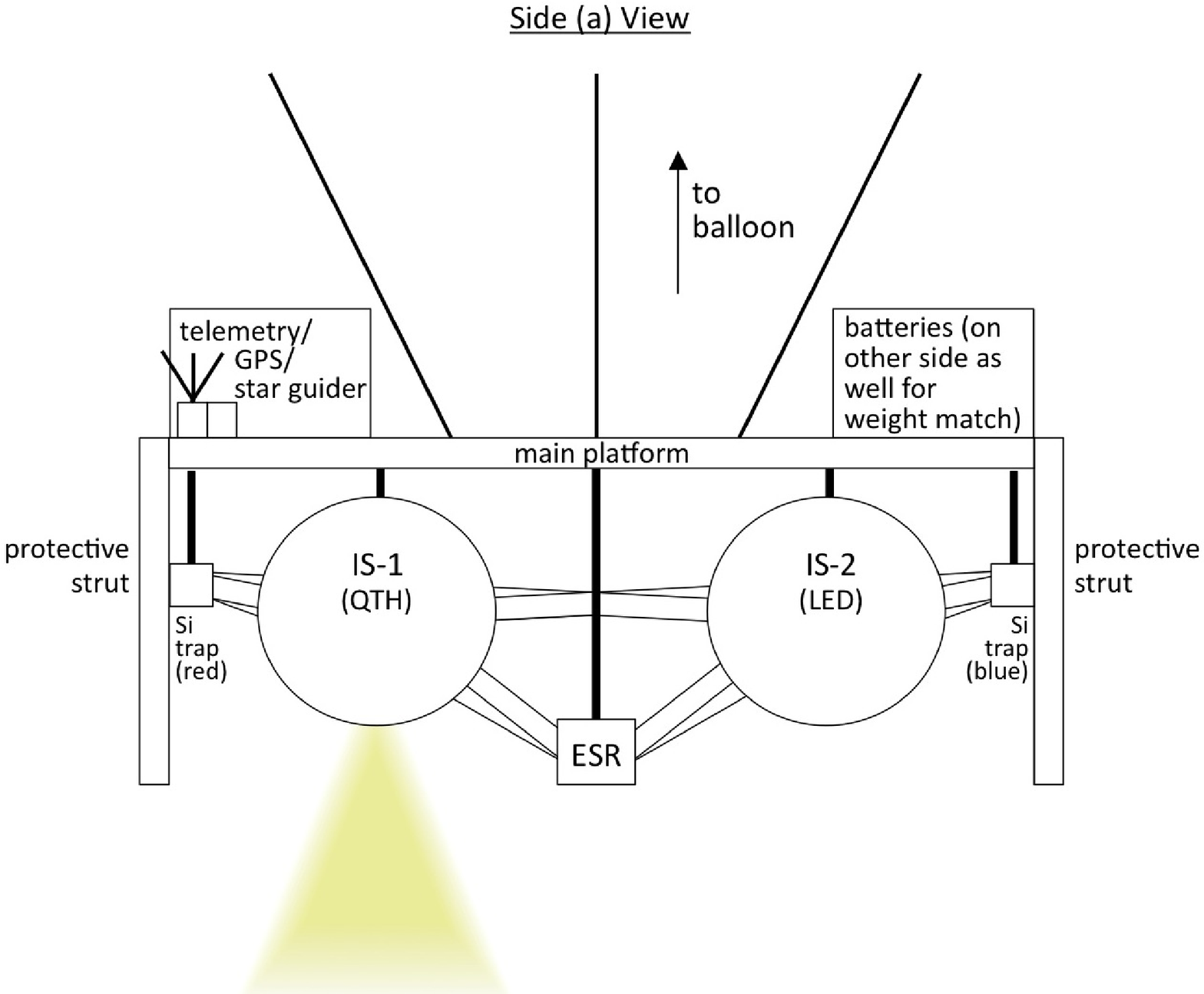}
\includegraphics[angle=0,width=7.7cm]{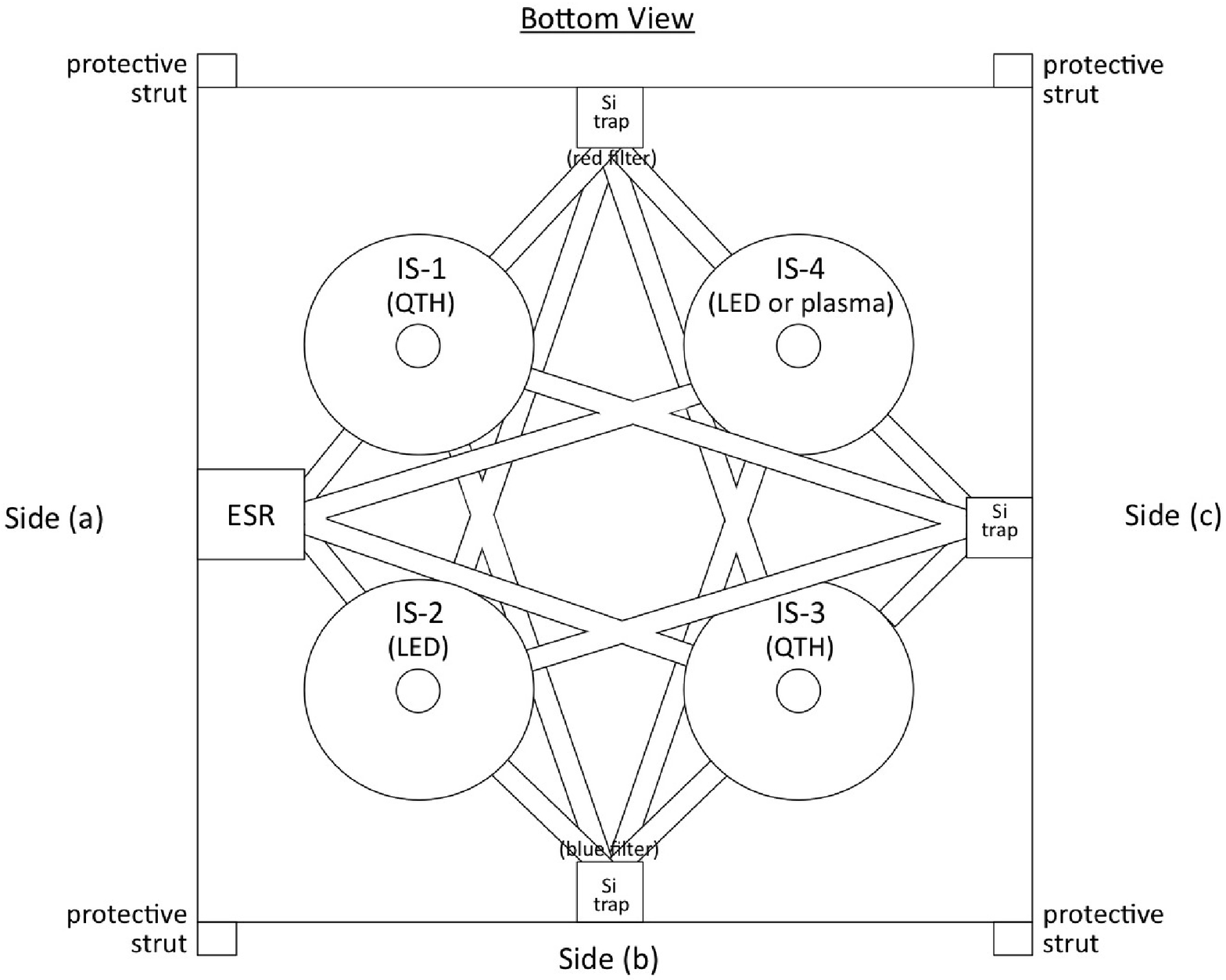}
\caption{
Conceptual sketch of a high altitude balloon-borne lamp calibration system for observatories.  The concept is based
on four integrating spheres, each sphere containing either a quartz-tungsten-halogen (QTH) lamp, or a light-emitting diode (LED) or
plasma lamp.  An electrical substitution radiometer (ESR), along with silicon trap photodiodes, provides absolute radiometry of
the sources.
}
\label{fig:PayloadSketch}
\end{center}
\end{figure}

Beam-based pulsed sources are not the only options for future photometric calibration standards above the atmosphere, however.  More isotropic,
and continuous, sources have the advantage that measurements are far less sensitive to the precise relative angle of the source and the observer,
and that one could time-average over atmospheric scintillation, rather than having to aperture-average.  Thus, a more isotropic calibrated source 
above the atmosphere is an attractive option.  No such sources presently exist.  In order to test such equipment for a future satellite of that type, 
and to obtain valuable data in the interim, the author and collaborators intend to fly calibrated light sources above observatories on a 
high-altitude balloon.   A very preliminary sketch of the payload we intend to fly may be seen in Fig.~\ref{fig:PayloadSketch}.  Such balloon flights
would not only provide important data for testing future satellite equipment, they would also have the advantage over satellite sources
that the payload may be recovered following the flight, and tested in the laboratory, rather than only having such laboratory tests preceeding the launch.
However, they of course cannot attain the altitude, nor can they easily attain the global reach, of a satellite.

One must note, however, that any future satellite-based light source would still clearly occupy a fixed orbit, and thus a rigorous program of atmospheric monitoring (using techniques such
as ground-based elastic LIDAR backscatter, and dual-band GPS for water vapor monitoring~[\citealt{GPS_H2O}], etc.)~would still be required to calibrate the full sky.
A device that could provide an ability to calibrate any selected point in the sky at chosen times, as well as
the features of visibility from major observatories and wavelength-tunability, is a far more distant concept, although recent progress in reusable suborbital vehicles~\citep{NASA_CRuSR}
could potentially point the way toward such an eventual possibility.  Even then, present and traditional means of atmospheric calibration
would likely remain necessary.  In any case, the addition of man-made calibrated light sources in space to the arsenal of techniques for 
photometric calibration will, however, provide a powerful new tool for increasing precision in astrophysics.

\section{CONCLUSION}
\label{conclusion}

In a campaign of observations of the 532 nm laser beam from the CALIPSO satellite, using a network of small cameras and photodiodes, we have
obtained measurements of atmospheric extinction at four dates and locations.  While the uncertainties are presently large, they are
dominated by sources that allow dramatic improvement, to the $\mathcal{O}(1\%)$ uncertainty level, with future data.  
Uncertainties due to atmospheric scintillation can be greatly
reduced by using optics with larger aperture.  
Beam profile-related uncertainties can be greatly reduced by precise pre-flight measurement.
A network of multiple 16'' telescopes, together with present and upcoming LIDAR satellites, will allow such improved datasets to be obtained.

Additionally, measurements of extinction using a more isotropic source are planned, which avoid the requirement of simultaneous precise measurement
of beam location as well as observed power.  Such equipment will initially be tested on high-altitude balloon flights.

Improved precision in photometric calibration will be nearly as critical for astronomy as increased aperture telescopes in upcoming decades.  The
future of precision photometry is extremely promising, and laboratory-based standards in space, such as described above, allow one to forsee
many-fold improvement in photometric calibration as a near-term prospect.

\acknowledgments

The author would like to acknowledge the work of the CALIPSO satellite team, led by PI Dr.~David M.~Winker and Project Scientist Dr.~Charles R.~Trepte
at NASA Langley Research Center (LaRC), and supported by NASA and CNES, as well as CALIPSO data obtained from the LaRC Atmospheric Science Data Center.
We would also like to acknowledge the critical help and advice from Dr.~Christopher Stubbs and Dr.~James Battat at Harvard University, 
Dr.~John McGraw and Dr.~Pete Zimmer at the University of New Mexico, Dr.~Susana Deustua at the Space Telescope Science Institute (STScI), Dr.~Chris 
Pritchet, Dr.~Robert Kowalewski, Dr.~Dean Karlen, Dr.~Russell Robb, Dr.~Julie Zhou, Mr.~Paul Scholz, Ms.~Sarah Maher, Ms.~Kristie Foster, Ms.~Grace Dupuis, Mr.~Kyle Fransham, 
Ms.~Kristin Koopmans, and Mr.~Michael Jarrett at the University of Victoria, and Dr.~Yorke Brown. 
The author was supported by the Canada Foundation for Innovation / British Columbia Knowledge 
and Development Fund Grant \#13075 and the University of Victoria.

\clearpage


\begin{thebibliography}{}

\bibitem[Albert \textit{et al.}, 2006]{alb06j} Albert, J., Burgett, W., and Rhodes, J.~2006. \textit{arXiv:astro-ph/0604339}.
\bibitem[Albrecht \textit{et al.}, 2006]{alb06} Albrecht, A., \textit{et al.}~2006. \textit{arXiv:astro-ph/0609591}.
\bibitem[Astier \textit{et al.}, 2006]{ast06} Astier, P., \textit{et al.}~2006. \aap~\textbf{441}, 31.
\bibitem[Bohlin, 2000]{boh00} Bohlin, R.C.~2000.  \aj~\textbf{120}, 437.
\bibitem[Bohlin and Gilliland, 2004]{boh04} Bohlin, R.C.~and Gilliland, R.L.~2004.  \aj~\textbf{127}, 3508.
\bibitem[Brun and Rademakers, 1997]{ROOT} Brun, R.~and Rademakers, F.~1997. Nucl.~Instr.~\& Meth.~in Phys.~Res.~A~\textbf{389}, 81.  See also \textit{http://root.cern.ch}.
\bibitem[Connolly \textit{et al.}, 2006]{con06} Connolly, A., \textit{et al.}~2006.\\ 
\textit{http://www.lsst.org/Science/Phot-z-plan.pdf}.
\bibitem[Eisenstein \textit{et al.}, 2001]{eis01} Eisenstein, D., \textit{et al.}~2001. \aj~\textbf{122}, 2267.
\bibitem[Environment Canada, 2010]{CanTempData} Environment Canada, National Climate Data and Information Archive~2010.\\
\textit{http://climate.weatheroffice.gc.ca/climateData/canada\_e.html}.
\bibitem[Fried, 1967]{fri67} Fried, D., 1967.  J.~Opt.~Soc.~Am.~\textbf{57}, 169.
\bibitem[Garber \textit{et al.}, 2007]{gar07} Garber, D.P., \textit{et al.}~2007.\\
\textit{http://www-angler.larc.nasa.gov/predict/}.
\bibitem[Healey and Kondepudy, 1994]{hea94} Healey, G.~and Kondepudy, R.~1994.  IEEE Trans.~Patt.~Anal.~Machine Intell.~\textbf{16}, 267.
\bibitem[Hufnagel, 1974]{huf74} Hufnagel, R.E.~1974. Optical Propagation Through Turbulence, OSA Technical Digest Series, paper WA1 
(Opt.~Soc.~Am., Washington, D.C.).
\bibitem[Hunt \textit{et al.}, 2009]{hun09} Hunt, W.H., \textit{et al.}~2009.  J.~Atmos.~Oceanic Technol.~\textbf{26}, 1214.
\bibitem[Jakeman \textit{et al.}, 1978]{jak78} Jakeman, E., \textit{et al.}~1978.  Contemp.~Phys.~\textbf{19}, 127.
\bibitem[Kaiser \textit{et al.}, 2008]{kai08} Kaiser, M.E., \textit{et al.}~2008.  Volume 7014 of \textit{SPIE}.
\bibitem[Kent \textit{et al.}, 2009]{ken09} Kent, S., \textit{et al.}~2009. \textit{arXiv:0903.2799}.
\bibitem[Knop \textit{et al.}, 2003]{kno03} Knop, R., \textit{et al.}~2003. \apj~\textbf{598}, 102.
\bibitem[Kodama \textit{et al.}, 1990]{kod90} Kodama, S., \textit{et al.}~1990. IEEE Trans.~Inst. Meas.~\textbf{39}, 230.
\bibitem[Kopp \textit{et al.}, 2005]{kop05} Kopp, G., Lawrence, G., and Rottman, G.~2005.  Solar Physics~\textbf{230}, 129.
\bibitem[Koester \textit{et al.}, 2007]{koe07} Koester, B., \textit{et al.}~2007. \apj~\textbf{660}, 221.
\bibitem[Minott, 1974]{min74} Minott, P.O.~1974.  NASA TM-X-723-74-122 (GSFC).
\bibitem[NAS Committee on Earth Science, 2008]{NAS_ACE} National Academy of Science, Committee on Earth Science and Applications
from Space~2008.  \textit{Satellite Observations to Benefit Science and Society: Recommended Missions for the Next Decade}
(National Academies Press, Washington, D.C.), p.~10.
\bibitem[NASA, 2010]{NASA_CRuSR} NASA Commercial Reusable Suborbital Research (CRuSR) program~2010.\\
\textit{http://crusr.arc.nasa.gov/}.
\bibitem[Nithianandam \textit{et al.}, 1993]{nit93} Nithianandam, J., Guenther, B.W., and Allison, L.J.~1993.  
Metrologia~\textbf{30}, 207.
\bibitem[NOAA, 2010]{USTempData} NOAA (National Oceanic and Atmospheric Administration), National Climatic Data Center~2010.\\
\textit{http://www.ncdc.noaa.gov/oa/ncdc.html}.
\bibitem[Pavlovsky \textit{et al.}, 2001]{pav01} Pavlovsky, C., \textit{et al.}~2001.  \textit{The ACS Instrument Handbook}, ver.~2.1 
(STScI, Baltimore).
\bibitem[Price \textit{et al.}, 2004]{pri04} Price, S.D., Paxson, C., and Murdock, T.L.~2004.  
Bull.~Amer.~Astron.~Soc.~\textbf{36}, 1457.
\bibitem[Roggemann and Welsh, 1996]{rog96} Roggemann, M.C.~and Welsh, B.~1996. \textit{Imaging Through Turbulence} (CRC Press, Boston), 
p.~62.
\bibitem[Savtchenko \textit{et al.}, 2008]{sav08} Savtchenko, A., \textit{et al.}~2008.  IEEE 
Trans.~Geosci. Remote Sens.~\textbf{46}, 2788.
\bibitem[Seas \textit{et al.}, 2007]{sea07} Seas, A., \textit{et al.}~2007.  Proc.~of \textit{CLEO} 2007, 
\textit{10.1109/CLEO.2007.4452342}.
\bibitem[Stubbs \textit{et al.}, 2006]{stu06} Stubbs, C., \textit{et al.}~2006. \apj~\textbf{646}, 1436.
\bibitem[Tatarski, 1961]{tat61} Tatarski, V.I.~1961.  \textit{Wave Propagation in a Turbulent Medium} (McGraw-Hill, New York), p.~169, 238.
\bibitem[Tregoning \textit{et al.}, 1998]{GPS_H2O} Tregoning, P., \textit{et al.}~1998.  JGR~\textbf{103}, 701.
\bibitem[Winker \textit{et al.}, 2009]{win09} Winker, D.M., \textit{et al.}~2009.  J.~Atmos. Oceanic Technol.~\textbf{26}, 2310.
\bibitem[Wood-Vasey \textit{et al.}, 2007]{woo07} Wood-Vasey, W.M., \textit{et al.}~2007. \apj~\textbf{666}, 694.
\end{thebibliography}
\end{document}